\documentclass[fleqn,usenatbib]{mnras}

\usepackage{mathptmx}
\usepackage{txfonts}
\usepackage[T1]{fontenc}
\usepackage{ae,aecompl}
\usepackage{pdflscape}	
\usepackage{afterpage}
\usepackage{times}
\usepackage{lscape}
\usepackage{longtable}
\usepackage{float}
\usepackage[normalem]{ulem}

\pagerange{\pageref{firstpage}--\pageref{lastpage}}

 \usepackage{graphicx}	

\makeatletter
\newcommand\footnoteref[1]{\protected@xdef\@thefnmark{\ref{#1}}\@footnotemark}
\makeatother
\def\ltsim{\mathrel{\hbox{\rlap{\hbox{\lower3pt\hbox{$\sim$}}}\hbox{$<$}}}}
\def\gtsim{\mathrel{\hbox{\rlap{\hbox{\lower3pt\hbox{$\sim$}}}\hbox{$>$}}}}
\def\p{$\pm$}
\def\Msun{M$_{\odot}$}
\def\mbh{$M_{\rm BH}$}
\def\Rsun{$R_{\odot}$}
\def\zsun{$z_{\odot}$}
\def\vpec{$\upsilon_{\rm pec}$}

\title[Gaia DR2 Astrometric Analysis of Black Hole X-ray Binaries]{{\em Gaia} DR2 Distances and Peculiar Velocities for Galactic Black Hole Transients}

 \author[Gandhi et al. 2018]{Poshak Gandhi,$^{1}$\thanks{E-mail: p.gandhi@soton.ac.uk}
 Anjali Rao,$^1$
 Michael A.C. Johnson,$^{1,2}$
 John A. Paice,$^{1,3}$\newauthor
 Thomas J. Maccarone$^4$
\\
 $^{1}$Department of Physics \& Astronomy, University of Southampton, Highfield, Southampton SO17 1BJ, UK\\
 $^{2}$Electronics and Computer Science, University of Southampton, Highfield, Southampton, SO17 1BJ, UK\\
 $^{3}$Inter-University Centre for Astronomy and Astrophysics, Pune, Maharashtra 411007, India\\
 $^{4}$Department of Physics \& Astronomy, Box 41051, Science Building, Texas Tech University, Lubbock, TX 79409-1051, USA\\
}

 \date{Accepted 2019 Feb 05.}
 \pubyear{2019}
\begin{document}
 \label{firstpage}

\maketitle
\begin{abstract}
We report on a first census of Galactic black hole X-ray binary (BHXRB) properties with the second data release (DR2) of {\em Gaia}, focusing on dynamically confirmed and strong candidate black hole transients. DR2 provides five-parameter astrometric solutions including position, parallax and proper motion for 11 of a sample of 24 systems. 
Distance estimates are tested with parallax inversion as well as Bayesian inference. We derive an empirically motivated characteristic scale length of $L$\,=\,2.17\,\p\,0.12\,kpc for this BHXRB population to infer distances based upon an exponentially decreasing space density prior. 
     Geometric DR2 parallaxes provide new, independent distance estimates, 
     but the faintness of this population in quiescence results in relatively large fractional distance uncertainties. Despite this, DR2 estimates generally agree with literature distances. The most discrepant case is BW\,Cir, for which detailed studies of the donor star have suggested a distant location at $\gtsim$\,25\,kpc. A large DR2 measured parallax and relatively high proper motion instead prefer significantly smaller distances, suggesting that the source may instead be amongst the nearest of XRBs. 
     However, both distances create problems for interpretation of the source, and follow-up data are required to resolve its true nature. 
DR2 also provides a first distance estimate to one source, MAXI\,J1820+070, and novel proper motion estimates for 7 sources. Peculiar velocities relative to Galactic rotation exceed $\approx$\,50\,km\,s$^{-1}$ for the bulk of the sample, with a median system kinetic energy of peculiar motion of $\sim$\,5\,$\times$\,10$^{47}$\,erg. BW\,Cir could be a new high-velocity BHXRB if its astrometry is confirmed. A putative anti-correlation between peculiar velocity and black hole mass is found, as expected in mass-dependent BH kick formation channels, but this trend remains weak in the DR2 data.
 \end{abstract}

%
%
\section{Introduction}

Black hole X-ray binaries (BHXRBs) are key sources for studying accretion in the strong gravitational regime. About 60 low mass BHXRBs are known in the Galaxy, though only about 20 of these are well characterised and confirmed as hosting black holes, based upon dynamical measurements of the companion star orbiting around the primary \citep{casares14, blackcat, watchdog}.

Distances to most XRBs remain highly uncertain. The most reliable distances come from geometric estimates based upon measurements of trigonometric parallax shifts relative to background sources. However, such measurements require exquisitely sensitive observations over several years, which have not been generally available except in the radio for a handful of cases where very long baseline interferometric studies have been possible: e.g. Cyg\,X--1 \citep{reid11}, V404\,Cyg \citep{millerjones09} and GRS\,1915+105 \citep{reid14_grs1915}. In the absence of geometric distances, estimates become reliant upon emission- and evolutionary-dependent model assumptions (cf., review by \citealt{jonker04}). Testing the validity of such model-dependent distances is important in order to properly model source luminosities and understand
the origin and evolution of these systems.

The {\em Gaia} mission \citep{gaiamission} has enabled a step change in the field of astrometric measurements of stars within the Galaxy. Geometric parallaxes 
along with highly accurate positions and proper motions for more than 1 billion stars are expected \citep{gaiamission}. Among the population of Galactic sources, BHXRBs are expected to be found. The typical brightness of these systems in quiescence is close to $V$\,$\sim$\,20\,mag or fainter. {\em Gaia} is expected to probe sources down to $G$\,$\approx$\,21\,mag, where the $G$ band covers 330--1050\,nm. So the viability and utility of astrometric measurements with {\em Gaia} is not clear {\em a priori}. 

A significant new data release (DR2, \citealt{gaiadr2}) occurred on 2018 April 25, and we present here an analysis of the properties of Galactic BHXRBs as seen in DR2, focusing on systems that are well characterised in the optical -- they are all either known to host a black hole based upon orbital dynamical measurements of the donor star, or are thought to be strong candidates to lie in this class. DR2 covers data gathered over 22 months from July 2014 to May 2016. We present astrometric measurements, including parallax and proper motion measurements where available, and examine their quality. These measurements are used to infer parallax-based distance estimates, and are compared with estimates from the literature based upon non-geometric techniques. We also present estimates of the peculiar Galactic velocities of the sources based upon three-dimensional kinematics, and carry out tests aimed at ascertaining formation pathways. This work is meant as a first look at the census of BHXRBs in this massive new parameter space of optical astrometric measurements. Future data releases can be expected to further improve upon this. 

\section{Gaia Counterpart Search and Association}

We queried the DR2 public release on 2018 April 25. A search radius of 2\,$\arcsec$ was used for DR2 entries around each of 24 BHXRBs, though we also experimented with larger radii in individual ambiguous cases as detailed later. The targets are chosen from the BlackCAT catalogue \citep{blackcat} as having dynamically confirmed mass measurements or other strong evidence of hosting a black hole. In addition, we also included a new, bright BHXRB transient candidate MAXI\,J1820+070 \citet{kawamuro18}. Finally, our list includes Cygnus\,X--1. Though a fairly persistent high mass X-ray binary, this source is amongst the best studied of black hole systems, and serves as a useful comparison source in terms of demonstrating the potential of {\em Gaia} for X-ray binary studies at the bright end. In order to assign {\em Gaia} counterparts, we used coordinates from the literature which are based upon optical, infrared or radio identifications. We stress that we did not use primary source coordinates listed in the {\sc simbad} database\footnote{\url{http://simbad.u-strasbg.fr/simbad}}, as {\em Gaia} associations therein could be spurious; this point is discussed further below.

In Table\,\ref{tab:sample}, we list all BHXRBs together with the associated {\em Gaia} counterpart. Literature J2000 coordinates are listed together with the {\em Gaia} coordinates for epoch 2015.5 and the counterpart $G$ mag, where available. No conversion between epochs is carried out here using the DR2 proper motions discussed later for some sources. We present the {\em Gaia} coordinates as reported in DR2 for ease of comparison with the archive. Maximal coordinate shifts expected based upon the proper motions are $\ltsim$\,0\farcs 15 between the two epochs, small enough not to affect the counterpart identification discussed here. 

\section{Distance Estimation with \textsl{Gaia}}
\label{sec:priors}

For a measured geometric parallax $\pi$ in milliarcsec (mas), distance can be obtained by parallax-inversion as $r_{\rm inv}$\,=\,$\frac{1}{\pi}$\,kpc. 
However, parallax inversion serves as a poor estimator, especially when uncertainties on $\pi$ are large ($\gtsim$\,20\,\%; cf. \citealt{bailerjones2015}).
This becomes increasingly relevant in the faint regime 
for the quiescent BHXRBs studied here, which tend to have magnitudes of $G$\,$>$\,19. 

Treating the problem as a case of Bayesian inference can help improve the estimate, if an appropriate choice of prior distribution of expected distances is possible. Several works have attempted various priors for Galactic {\em stellar} populations (e.g., \citealt{bailerjones2015}, \citealt{astraatmadja16a}, \citealt{astraatmadja16b}, \citealt{bailerjones2018}). But the validity of stellar-based priors is uncertain when it comes to the BHXRB population for two reasons. Firstly, there is the possibility of natal kicks scrambling the spatial distribution of XRBs relative to the underlying stellar population. Secondly, the selection of these sources differs markedly from stars: they are discovered in X-rays, which have a longer mean free path through Galactic gas and dust than the optical. Discovered sources are then often subject to targetted multiwavelength follow-up, which is typically much more sensitive than field stars. Intuitively, this would argue for a larger characteristic scale length for BHXRBs.

  In the following section, we set up two possible priors for our sample: a simple empirical exponential model based upon the known BHXRB population, as well as a three-dimensional model of the space distribution of XRBs in X-rays. 
  
The prior $P(r)$ is combined with a likelihood function $P(\pi|r)$ to produce the final posterior estimate of $r$, $P(r|\pi)$. 
The likelihood $P(\pi|r)$ is assumed to be normally distributed with mean of $\frac{1}{r}$ and standard deviation $\sigma_\pi$, as described in the above works. 
We quote the mode ($r_{\rm mode}$) of the posterior as the final distance estimate and the uncertainties represent the highest density confidence interval, corresponding to a 1-$\sigma$ interval for a Gaussian distribution. As described in \citet{bailerjones2018}, this is equivalent to lowering down a horizontal line from $r_{\rm mode}$, until the enclosed probability encompasses 68.3\,\% of the full posterior probability density. 

We have not considered systematic uncertainties in our calculation, as recommended in \citet{gaiadr1}. Systematic astrometric uncertainties have not been accounted for individually in DR2, though generic corrections are included (\citealt{gaiadr2}; \citealt{lindegren18}). Residual zeropoint systematics are expected to be only $\approx$\,0.03\,mas, much smaller than typical statistical uncertainties in the faint regime (Ibid.). Global systematics may be as large as 0.1\,mas, but the Collaboration recommendation at present is to not combine these with statistical uncertainties, but rather keep them in mind for the interpretation (Ibid.).

Table\,\ref{tab:sourceswithastrometry} summarises astrometric parameters for the sources where available. Values of parallax and proper motion are stated, together with several DR2 pipeline flags relevant to the fits. These flags are described in the {\em Gaia} Data Model\footnote{\label{note:dr2model}{\em Gaia} Data model:\newline \url{https://gea.esac.esa.int/archive/documentation/GDR2/Gaia_archive/chap_datamodel}}, and will be discussed in detail in Section\,\ref{sec:discussion}. 

\section{Distance Priors for the BHXRB population}

\subsection{Exponential Model}
\label{sec:expprior}

  This prior adopts an exponentially decreasing space density with distance $r$ from the Sun, as in several recent works focusing on distributions of stars in the Galaxy (\citealt{astraatmadja16a}, \citealt{astraatmadja16b}, \citealt{bailerjones2018}). However, we determine a scale length more appropriate to dynamically confirmed BHXRBs as follows.

  We first collated previous estimates ($r_{\rm lit}$) of distances to our BHXRB sample from the literature. These distances come from a number of published works, and are based upon a variety of techniques (sometimes used in combination) including photometric and spectral classification of the companion star, radio parallax measurements, and constraints based upon line-of-sight reddening columns. These $r_{\rm lit}$ estimates, together with uncertainties, are listed in Table\,\ref{tab:results}. We then fitted these distances with the following exponential prior probability model: 

  \begin{equation}
    P(r) = \frac{1}{2L^3} r^2 e^{-r/L}~~~~~{\rm for}~~ r>0, 
  \end{equation}

\noindent
where $L$ is the scale length and the shell volume element component scales with $r^2$. Such a procedure is very similar to the position-dependent scale length fitting carried out by \citet{bailerjones2018} for the Galactic stellar population, albeit with a small sample size here. 

An unbinned maximum likelihood algorithm\footnote{\url{https://www.harrisgeospatial.com/docs/ml\_distfit.html}} was used for the fit. In order to determine the uncertainty on $L$, randomised ensembles of $r_{\rm lit}$ values were generated by resampling from a Normal distribution for each object, with the assumed mean and standard deviation being the published value of $r_{\rm lit}$ and its uncertainty, respectively. There are four sources with lower limits on $r_{\rm lit}$, for which we drew random values uniformly between $r_{\rm lit}$ and $r_{\rm lit}$\,+\,5\,kpc, a threshold based upon typical upper limits suggested in the literature. We also tested that our fits are not strongly sensitive to this threshold; e.g. doubling the threshold to $r_{\rm lit}$\,+\,8\,kpc changes our fitted $L$ by less than the scatter quoted below. 

\begin{figure*}
 \centering
     \includegraphics[width=85mm]{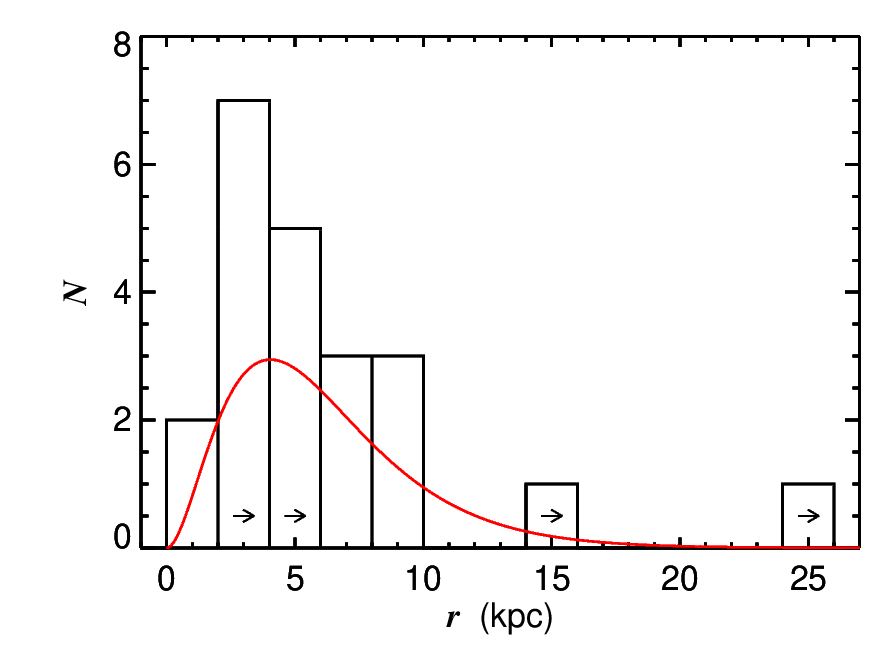} 
     \includegraphics[width=85mm]{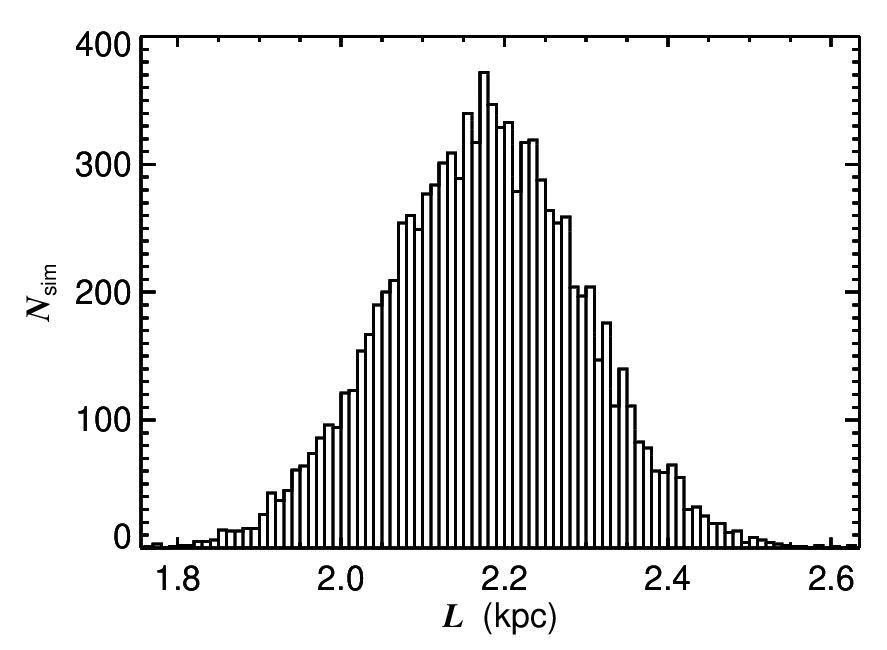} 
     \caption{Fitting of a characteristic scale length $L$ for the spatial distribution of known BHXRBs. $r$ represents distance from the Sun. The panel on the left shows the best fitting model to the sample in red, and the panel on the right shows the histogram of $L$ values resulting from the resampling simulations. See section\,\ref{sec:expprior} for details.}
 \label{fig:expmodel}    
 \end{figure*}

A total of 10,000 ensembles were randomised, resulting in a measured mean value of the characteristic scale length $L$\,=\,2.17\,\p\,0.12\,kpc, with the uncertainty here being the standard deviation amongst the randomised ensembles. The prior model and the distribution of randomised scale lengths are plotted in Fig.\,\ref{fig:expmodel}. The posterior estimator of the distance based upon this prior is denoted hereafter as $r_{\rm exp}$. 

\subsection{Milky Way X-ray source distribution}

Our second tested prior is based upon X-ray detections of XRBs in the Milky Way. Our model is taken from \citet{grimm02}, who present several equations to describe the spatial density of sources separately in the Galactic thin disc, bulge and halo. Their base model, in turn, was built from early works describing the Galactic stellar mass distribution \citep{bahcallsoneira80, dehnenbinney98}. The space density of BHXRBs is taken to correspond to the Galactic low mass X-ray binary (LMXB) density as described by Eqs.\,(4), (5) and (6) of \citet{grimm02}, representing the density in the bulge, disk, and halo of the Milky Way, respectively.

  This is a three-dimensional model with a scale length depending upon Galactic coordinates. The full equations and relevant constants for the model can be found in \citet{grimm02}, and are also reproduced in the Appendix. 
The number density prior, $P(r)$, is composed of the space density multiplied by $r^2$ representing the volume element, with $r$ representing distance from the Sun, as for the exponentially-decreasing space density prior. The posterior estimate of the distance is denoted by $r_{\rm MW}$. 
~\par
~\par
\noindent
In Sections\,\ref{sec:results} and \ref{sec:discussion}, we will compare and discuss these various distance estimates.

\section{Kinematics}
\label{sec:vpecintro}

The space velocities of XRBs remain poorly known. If black holes originate as remnants of supernova explosions, they may receive natal kicks that could potentially recoil and scatter them into new trajectories. Symmetric as well as asymmetric natal kicks have been debated for binary systems \citep{blaauw61,brandtpodsiadlowski95}. Studying the distribution of BHXRB locations and their kinematics in excess of Galactic rotation can thus shed light on their origins, and there are various works that have attempted this in the past \citep{vanparadijswhite95, jonker04, repetto17}.

Two black hole binaries for which there is evidence of strong dynamical scattering are XTE\,J1118+480 and GRO\,J1655-40 \citep{mirabel17review}. But such measurements typically require full astrometric and kinematic information: i.e. parallaxes ($\pi$), proper motions ($\mu_{\alpha}$, $\mu_{\delta}$ in RA, Dec) and radial velocities ($\gamma$), and thus exist for only a handful of sources which have sensitive high angular resolution observations spread over time. The best of these so far come from radio very long baseline interferometric observations. 

  DR2 now provides measurements of ($\mu_{\alpha}$\,cos$\delta$, $\mu_{\delta}$) in addition to $\pi$, in the optical. Together with previous measured values or assumed plausible ranges of $\gamma$, this allows us to estimate source velocities. We used the formalism of \citet{reid09} to transform the observables into heliocentric space velocities, and then remove Solar motion as well as Galactic rotation to determine peculiar velocities \vpec\ (i.e. relative to expected motion in the Galaxy). 
  Relevant constants are listed in Table\,\ref{tab:constants}. Uncertainties were determined using Monte Carlo resampling of all parameters including the constants.

  Whereas literature distance estimates are available for many of our BHXRBs, the DR2 proper motion measurements are novel for most. Furthermore, as we discuss below, typical uncertainties on these proper motions are small. So one would expect to be able to infer interesting dynamical information for most sources. However, this is hampered by the large uncertainties on $\pi$. In order to mitigate these uncertainties, for the systems with particularly poor DR2 parallax measurements $|\frac{\sigma_\pi}{\pi}|$\,$>$\,1, we chose to use the literature distance estimates, $r_{\rm lit}$, instead of the {\em Gaia} distance estimates, when inferring \vpec\ since the uncertainties on $r_{\rm lit}$ are typically smaller. This affected only 3 sources (XTE\,J1118+480, 4U\,1543--475, Swift\,J1753.5--0127), and allows us to better focus on the novel kinematic information provided by DR2 in these cases. We verified that using the full DR2 astrometry also resulted in consistent results in these cases.

\section{Results}
\label{sec:results}

DR1 reported detections and $G$ mags of several BHXRBs in our sample, but without parallaxes or proper motion estimates. DR2 is more complete is terms of sky coverage and depth. 

We find potential DR2 counterparts within 2\,\arcsec of the known positions of 21 BHXRBs, listed in Table\,\ref{tab:sample}. Examining these fields individually, we find that in 18 cases, the identified {\em Gaia} counterparts can be associated with the BHXRB. Two sources, GX\,339--4 and V404\,Cyg, had multiple possible counterparts within 2\,\arcsec, and the final association was confirmed from close examination of the fields. Finding charts for both cases are presented in the Appendix. The offset from the literature position is found to be less than 1\,$\arcsec$ in most cases, with only one secure counterpart (for V4641\,Sgr) lying at an offset of just above 1\,\arcsec.

All shortlisted counterparts at larger separations are, in fact, unassociated with the parent BHXRB. Of these, we note a potential confusion issue for the particular case of GRS\,1009--45. The nearest DR2 neighbour is a star with DR2 designation 5414249796406957056 lying about 1\farcs 2 to the south--east of the BlackCAT position, with DR2 coordinates of RA\,=\,10:13:36.40, Dec\,=\,--45:04:32.52 (epoch 2015.5). The BHXRB itself is fainter in the optical, and is not detected in DR2. A finding chart is presented in the Appendix. We highlight this source because at the time of writing this manuscript, the {\sc simbad} database has associated the above (incorrect) neighbouring star with the BHXRB.\footnote{\url{http://simbad.u-strasbg.fr/simbad/sim-id?Ident=GRS+1009-45}} We stress that this association is spurious.

Complete DR2 astrometric solutions consisting of position, parallax and proper motion are available for 11 of the 18 associated systems (Table\,\ref{tab:sourceswithastrometry}). 
But uncertainties $\sigma_\pi$ on parallax in most cases are significant, with a median $\pi/\sigma_\pi$\,=\,3.2 and a fractional uncertainty $>$\,1 for 3 sources (XTE\,J1118+480, 4U\,1543--475 and Swift\,J1753.5--0127). The reported parallax in one case is negative (Swift\,J1753.5--0127). Negative parallaxes can arise in cases of high fractional uncertainty, as a result of astrometric fits to noisy data \cite[e.g. ][]{bailerjones2015}. The weak parallax constraints are not surprising when considering the faintness of the detected DR2 counterparts. The median mag of our sample is $G$\,=\,19.35 mag, with a standard deviation of 3.18\,mag. 

\begin{figure}
 \centering
 \includegraphics[width=85mm,angle=0]{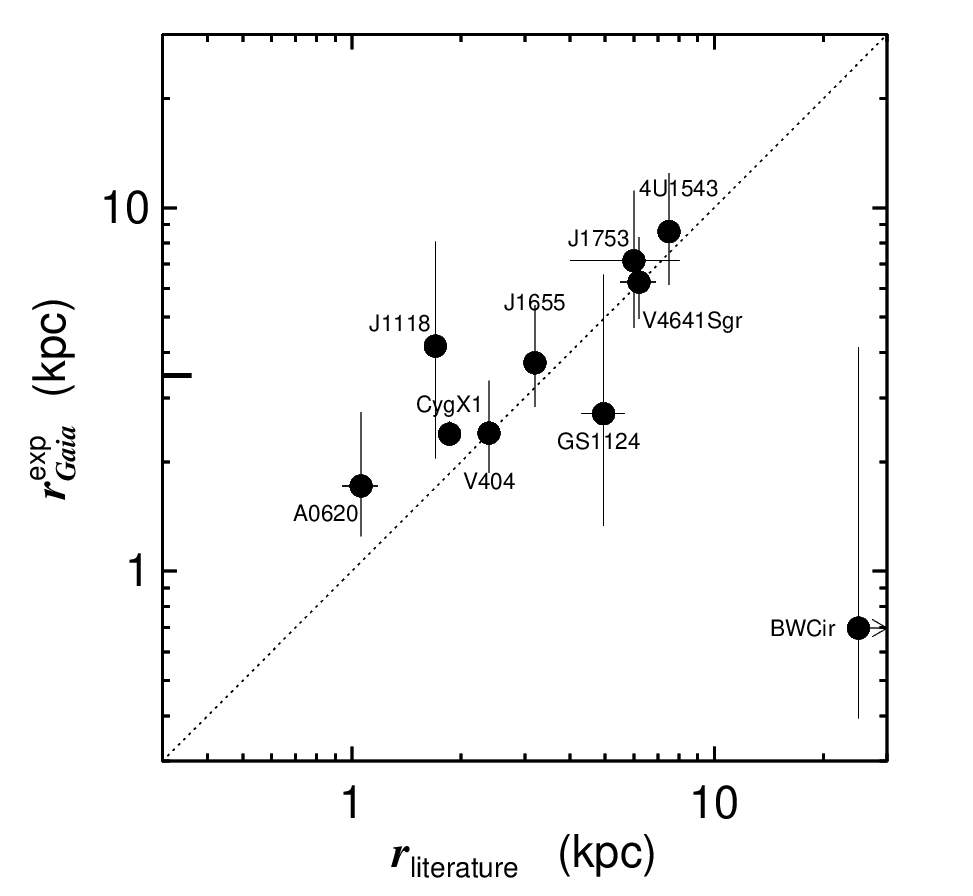} 
 \caption{Comparison of {\em Gaia} DR2 distances (based on the expoential prior, $r_{\rm exp}$) with literature estimates ($r_{\rm lit}$) for our BHXRB targets, where known. The dotted line represents equality between the two. 
   The long dash at $r_{\rm exp}$\,$\approx$\,3.5\,kpc denotes the most likely value of $r_{\rm exp}$ for MAXI\,J1820+070, a new XRB that lacks $r_{\rm lit}$ estimates.}
 \label{fig:distancecomparison}
 \end{figure}

Distance estimates and 1-$\sigma$ confidence intervals based upon the DR2 parallaxes are reported in Table\,\ref{tab:results}. We report $r_{\rm inv}$, $r_{\rm exp}$, and $r_{\rm MW}$, as described in Section\,3. There is no case of strong disagreement between these within the uncertainties, and the Bayesian priors also yield positive distance estimates for the one case with negative $\pi$. 

  In Fig.\,\ref{fig:distancecomparison}, we compare one of these estimates ($r_{\rm exp}$) with the literature distances $r_{\rm lit}$ compiled in Table\,\ref{tab:results}, for the 10 systems where both are available. The two kinds of estimates agree within uncertainty for 6 objects. There is only modest discrepancy in two cases: A\,0620--00 and XTE\,J1118+480, which is not surprising given the high fractional $\pi$ uncertainties. Two cases disagree more strongly: BW\,Cir and Cyg\,X--1. We will delve into individual systems in the Discussion section. 

\begin{figure*}
 \centering
     \includegraphics[width=150mm]{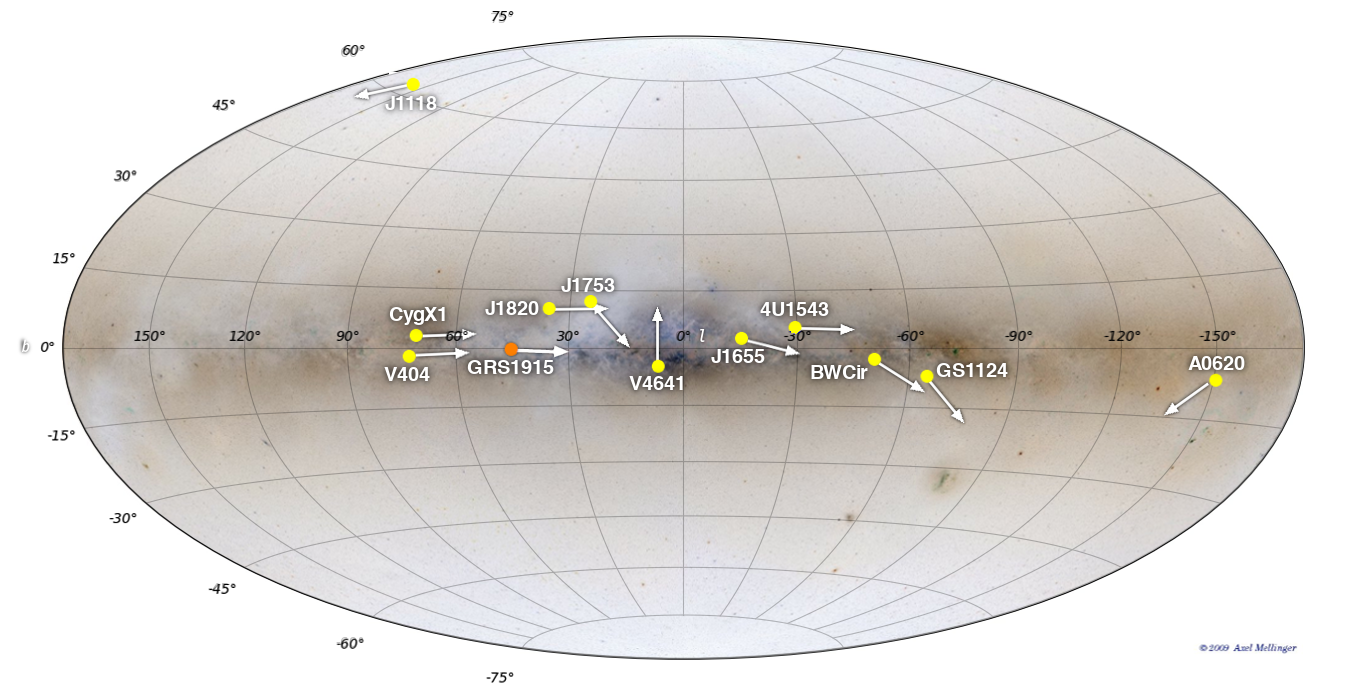} 
     \caption{Schematic illustration of the sky-projected motions of BHXRBs in the Milky Way. The layout is in Galactic Coordinates, using the Hammer-Aitoff projection. Yellow circles represent the locations of BHXRBs for which we have proper motions from {\em Gaia} DR2, while the orange circle represents radio measurements for GRS\,1915+105 (\citealt{dhawan07}, \citealt{reid14_grs1915}).
       The white arrows show the direction of their proper motion. (Galactic sky image: \citealt{mellinger09}).}
 \label{fig:hammer} 
 \end{figure*}

DR2 proper motion measurements for the 11 systems with astrometric solutions are typically much more constrained than their parallaxes, with fractional uncertainties of only $\approx$\,0.05. The mean magnitude of the proper motions is 7.3\,\p\,0.2\,mas\,yr$^{-1}$, most of them being in the direction of decreasing projected RA and also in decreasing Dec. The source with the highest magnitude of proper motion is XTE\,J1118+480, about 2.5 times the average value above. In Fig.\,\ref{fig:hammer}, the sources are plotted on a Hammer-Aitoff projection of the sky, along with arrows showing the current direction of their proper motion. In Galactic coordinates, there is no significant average motion in Galactic latitude ($b$) after excluding the outlier XTE\,J1118+480. There is, however, an overall average motion towards decreasing Galactic longitude $\mu_l$\,cos($b$)\,=\,--3.2\,\p\,0.2\,mas\,y$^{-1}$, also discernible in Fig.\,\ref{fig:hammer}, though we note the small size of this sample. Since these motions are heliocentric, Solar motion along the direction of Galactic rotation, combined with differential disc rotation, should contribute to this trend.
  
  Table\,\ref{tab:results} also lists the peculiar velocities relative to Galactic rotation. The median value of $\upsilon_{\rm pec}$ for these 11 systems is 80.9\,km\,s$^{-1}$, but there is also substantial scatter 
  resulting from the large parallax uncertainties in individual cases. Further discussion on the $\upsilon_{\rm pec}$ values and their implications will follow in the next section. 

  Distance and peculiar velocity posterior distributions are presented in the Appendix for all sources.

\begin{centering}
  \begin{table*}
    \caption{{\em Gaia} DR2 astrometric and photometric data for our sample of 24 Galactic black hole X-ray binaries.}
    \begin{tabular}{llccccccr}
      \hline
      \hline
      
      \#  & Source & \multicolumn{3}{c}{\underline{\hspace*{1.4cm}Literature\hspace*{1.4cm}}} & \multicolumn{2}{c}{\underline{\hspace*{1.1cm}{\em Gaia}\hspace*{1.1cm}}} & Offset & $G$ \\
          &        & (h:m:s) & (d:m:s) & Ref. & (h:m:s) & (d:m:s) & (arcsec) & mag \\
  (1) & (2) & (3) & (4) & (5) & (6) & (7) & (8) & (9) \\
      \hline
      1 & GRO\,J0422+32      & 04:21:42.79 & +32:54:27.10 & [1] & 04:21:42.72 & +32:54:26.94     &0.86 & 20.85 \\
      2 & A\,0620--00       & 06:22:44.50 & --00:20:44.72 & [2] & 06:22:44.54 &--00:20:44.37     &0.72 & 17.52 \\
      3 & GRS\,1009--45      & 10:13:36.34 &--45:04:31.50 & & \sout{10:13:36.40} &\sout{--45:04:32.52} &\sout{1.22} & \sout{17.73} \\
      4 & XTE\,J1118+480     & 11:18:10.79 & +48:02:12.42 & [3] & 11:18:10.77 & +48:02:12.21  &0.33 & 19.35 \\
      5 & GS\,1124--684      & 11:26:26.65 &--68:40:32.83 & & 11:26:26.59 &--68:40:32.90     &0.35 & 19.57 \\
      6 & Swift\,J1357.2--0933&13:57:16.82 &--09:32:38.55 & & 13:57:16.84 &--09:32:38.79     &0.34 & 20.75 \\
      7 & BW\,Cir            & 13:58:09.70 &--64:44:05.80 & [4] & 13:58:09.71 &--64:44:05.29     &0.51 & 20.53 \\
      8 & 4U\,1543--475      & 15:47:08.32 &--47:40:10.80 & & 15:47:08.27 &--47:40:10.37     &0.70 & 16.48 \\
      9 & XTE\,J1550--564    & 15:50:58.70 &--56:28:35.20 & [5] & 15:50:58.65 &--56:28:35.31     &0.41 & 20.90 \\
      10 & XTE\,J1650--500   & 16:50:00.98 &--49:57:43.60 & [6] & \sout{16:50:00.81} &\sout{--49:57:43.76}     &\sout{1.67} & \sout{20.02} \\
      11 & GRO\,J1655--40    & 16:54:00.14 &--39:50:44.90 & [7] & 16:54:00.14 &--39:50:44.88     &0.05 & 16.22 \\
      12 & MAXI\,J1659--152  & 16:59:01.68 &--15:15:28.73 & [8] &          -- &           --     & -- & -- \\
      13 & GX\,339--4        & 17:02:49.40 &--48:47:23.30 & [9] & 17:02:49.38 &--48:47:23.16     &0.23 & 16.47 \\
      14 & H\,1705-250 	     & 17:08:14.52 &--25:05:30.15 & [10] & 17:08:14.50 &--25:05:30.29     &0.30 & 20.83 \\
      15 & XTE\,J1752--223   & 17:52:15.10 &--22:20:32.36 & [11] & 17:52:15.11 &--22:20:31.43     &0.96 & 20.20 \\
      16 & Swift\,J1753--0127& 17:53:28.29 &--01:27:06.22 & [12] & 17:53:28.29 &--01:27:06.31     &0.09 & 16.70 \\
      17 & XTE\,J1817--330   & 18:17:43.53 &--33:01:07.47 & [13] & \sout{18:17:43.49} & \sout{--33:01:08.80} & \sout{1.41} & \sout{19.86} \\
      18 & V4641\,Sgr        & 18:19:21.58 &--25:24:25.10 & & 18:19:21.63 &--25:24:25.84     &1.04 & 13.57 \\
      19 & MAXI\,J1820+070   & 18:20:21.93 & +07:11:07.08 & [14] & 18:20:21.94 & +07:11:07.19&0.19  & 17.41 \\
      20 & XTE\,J1859+226    & 18:58:41.58 & +22:39:29.40 & [15] & -- & -- & -- & -- \\
      21 & GRS\,1915+105     & 19:15:11.55 & +10:56:44.76 & [16] & -- & -- & -- & -- \\
      22 & GS\,2000+251      & 20:02:49.48 & +25:14:11.36 & & 20:02:49.52 & +25:14:10.64     &0.92 & 21.22 \\
      23 & V404\,Cyg         & 20:24:03.82 & +33:52:01.90 & [17] & 20:24:03.82 & +33:52:01.84     &0.06 & 17.19 \\
      24 & Cyg\,X--1         & 19:58:21.67 & +35:12:05.73 & [18] & 19:58:21.67 &  +35:12:05.69  & 0.05 & 08.52 \\
      \hline
      \hline
    \end{tabular}\label{tab:sample}
    ~\par
    Notes: Cols. (3)--(5): ICRS coordinates (J2000.0) from the literature; Cols. (6)--(7): ICRS coordinates (J2015.5) of the nearest counterparts within 2\,\arcsec\, from {\em Gaia} DR2; (8) Offset in arcsec between the two sets of coordinates. Unassociated neighbours within 2\,\arcsec\ have been struck out. References for literature coordinates in Col. (5): Where not stated, the coordinates are taken from \citealt{blackcat}; [1] \citet{shrader94}; [2] \citet{gallo06}. [3] \citet{fender01_j1118}; [4] \citet{brocksopp01}; [5] \citet{corbel01_j1550}; [6] \citet{tomsick04_j1650}; [7] \citet{bailyn95}; [8] \citet{paragi13}; [9] \citet{corbel00}; [10] \citet{yang12_h1705}; [11] \citet{millerjones11_j1752}; [12] \citet{fender05_atel_j1753}; [13] \citet{rupen06_atel_j1817}; [14] \citet{kennea18}; [15] \citet{garnavich99_j1859}; [16] \citet{dhawan00}; [17] \citet{millerjones09}; [18] \citet{reid11}.
      \end{table*}
\end{centering}

\begin{table*}
  \caption{{\em Gaia} DR2 astrometric data for the sample of black hole binaries.}
    \begin{tabular}{lccccclccr}
  \hline
 \hline
   Source & Parallax  & \multicolumn{2}{c}{\underline{\hspace*{0.7cm}Proper Motion\hspace*{0.7cm}}} & \multicolumn{2}{c}{\underline{\hspace*{0.7cm}$\upsilon_{\rm radial}$\hspace*{0.7cm}}} & \multicolumn{4}{c}{\underline{\hspace*{2.5cm}Fit\hspace*{2.5cm}}}\\
    & $\pi$  & $\mu_{\alpha}$\,cos$\delta$ & $\mu_{\delta}$ & $\gamma$ & & {\tt n\_vis} & {\tt gof}& {\tt noise} & {\tt sigma5d}\\
    & (mas) & (mas y$^{-1}$) & (mas y$^{-1}$) & (km s$^{-1}$) & Ref. & & & (mas) & (mas) \\
  (1) & (2) & (3) & (4) & (5) & (6) & (7) & (8) & (9) &     (10) \\
 
 \hline

         A\,0620--00        &  0.64\,$\pm$\,0.16 & --0.09\,$\pm$\,0.25  & --5.20\,$\pm$\,0.30  & 8.5\,$\pm$\,1.8 & [1] & 10 & 3.026 & 0.52 [3.13] & 0.30 \\ 
         XTE\,J1118+480      &  0.30\,$\pm$\,0.40 & --17.57\,$\pm$\,0.34 & --6.98\,$\pm$\,0.43  & --15\,$\pm$\,10  & [2] & 15 & 4.694 & 1.05 [3.12] & 0.44\\ 
         GS\,1124--684       &  0.61\,$\pm$\,0.34 & --2.44\,$\pm$\,0.61  & --0.71\,$\pm$\,0.46  & 14.2\,$\pm$\,6.3  & [3] & 16 & 0.352 & 0.0 & 0.54 \\ 
         BW\,Cir        &   1.83\,$\pm$\,0.58  & --9.38\,$\pm$\,2.22 & --5.70\,$\pm$\,2.26 & 103\,$\pm$\,4  & [4] & 10 & 1.592 & 0.0 & 2.25\\ 
         4U\,1543--475       &   0.04\,$\pm$\,0.07  &   --7.41\,$\pm$\,0.14  &   --5.33\,$\pm$\,0.10  & --87\,$\pm$\,3  & [5] & 14 & 0.344 & 0.0 & 0.13 \\ 
        GRO\,J1655--40      & 0.27\,$\pm$\,0.08 & --4.20\,$\pm$\,0.13 & --7.44\,$\pm$\,0.09 & --142\,$\pm$\,1.5   & [6] & 12 & 4.035 & 0.20 [2.21] & 0.13 \\ 
        Swift\,J1753--0127  & --0.01\,$\pm$\,0.13 &  1.13\,$\pm$\,0.16  &  --3.53\,$\pm$\,0.15 & 6\,$\pm$\,6 & [7] & 10 & 0.570 & 0.0 & 0.18 \\ 
       V4641\,Sgr   & 0.15\,$\pm$\,0.04 & --0.73\,$\pm$\,0.07 & 0.42\,$\pm$\,0.06  & 107.4\,$\pm$\,2.9 & [8] & 9 & 0.675 & 0.06 [0.56] & 0.08 \\ 
        MAXI\,J1820+070     &   0.31\,$\pm$\,0.11  &   --3.14\,$\pm$\,0.19  &   --5.90\,$\pm$\,0.22  &   --  & -- & 10 & --0.871 & 0.0 & 0.25\\ 
        V404\,Cyg        &   0.44\,$\pm$\,0.10  &   --5.77\,$\pm$\,0.17  &   --7.85\,$\pm$\,0.17  & --0.4\,$\pm$\,2.2  & [9] & 15 & 9.525 & 0.50 [7.12] & 0.18 \\ 
        Cyg\,X--1           &   0.42\,$\pm$\,0.03  &   --3.88\,$\pm$\,0.05  &   --6.17\,$\pm$\,0.05  &         --5.1\,$\pm$\,0.5  & [10] & 16 & 3.626 & 0.0 & 0.04 \\ 
\hline
 \hline
\end{tabular}
    ~\par
    Notes: Col. (6) References for the radial velocity: [1] \citet{hernandez10}; [2] \citet{mirabel01}; [3] \citet{wu15_gs1124_paperI}; [4] \citet{casares04}; [5] \citet{orosz98}; [6] \citet{mirabel02}; [7] \citet{neustroev14}, though note that the authors stress potential caveats related to systematic uncertainties; [8] \citet{orosz01}; [9] \citet{casares94}; [10] \citet{gies08}. Col. (7): {\tt visibility\_periods\_used}: Number of distinct visibility periods used in the DR2 astrometric solution; Col. (8) {\tt astrometric\_gof\_al}: Gaussianised $\chi^2$ goodness-of-fit statistic; Col. (9) {\tt astrometric\_excess\_noise}: Excess residuals, expressed as an angle, of the data fit to the best astrometric model. Numbers in square brackets denote the corresponding significance of this value ({\tt astrometric\_excess\_noise\_sig}); Col. (10) {\tt astrometric\_sigma5d\_max}: The longest principal axis, expressed as an angle, in the 5-dimensional DR2 covariance matrix.
\label{tab:sourceswithastrometry}
\end{table*}
 

   \begin{center}
 \centering
 \begin{table*}
 \centering
  \caption{Distances and space velocity estimates based upon {\em \textit{Gaia}} astrometry.}
\medskip
    \begin{tabular}{l c c c c c r}
     \hline
     \hline
 Source & $r_{\rm inv}$ & $r_{\rm exp}$ & $r_{\rm MW}$ & \vpec & \multicolumn{2}{c}{\underline{\hspace*{0.95cm}$r_{\rm lit}$\hspace*{0.95cm}}}\\

   & (kpc) & (kpc) & (kpc) & (km s$^{-1}$) & (kpc) & Ref.\\
(1)   & (2) & (3) & (4) & (5) & (6) & (7) \\
     
\hline     

         GRO\,J0422+32  & -- & -- & -- & --  & $2.49\pm0.30$ & [1] \\
         A\,0620--00   &  1.57$\pm0.40$ & $ 1.72_{-0.47}^{+ 1.02}$ & 1.70$_{-0.44}^{+ 0.90}$ & 46.5$_{ -11.0}^{+ 38.0}$ & $1.06\pm0.12$ & [2] \\
         GRS\,1009--45 & -- & -- & -- & -- & $3.8\pm0.3$ & [3]  \\
         XTE\,J1118+480  &  3.34$\pm$4.42    & $4.16_{-2.12}^{+ 3.91}$ & 8.22$_{-6.71}^{+ 9.26}$ & 166.5$_{ -40.0}^{+ 47.0}\dag$ & $1.7\pm0.1$ & [4] \\
         GS\,1124--684  &   1.65$\pm0.93$ & $ 2.71_{-1.38}^{+ 3.82}$ & 4.01$_{-2.27}^{+ 3.91}$ & 74.5$_{ -27.0}^{+ 77.0}$ & $4.95_{-0.65}^{+0.69}$ & [5] \\ 
         Swift\,J1357.2--0933  & -- & --  & -- & -- & $>2.29$ & [6]  \\
         BW\,Cir  &  0.55$\pm$0.17 & $ 0.70_{-0.30}^{+ 3.43}$ & 7.01$_{-6.32}^{+ 3.43}$ &  114.5$_{ -17.0}^{+ 62.0}$ & $\gtsim$\,25 & [7] \\
         4U\,1543--475  &  24.72$\pm$41.15  & 8.60$_{-2.46}^{+ 3.84}$ & 9.47$_{-2.53}^{+ 3.91}$ & 93.9$_{ -11.6}^{+ 16.2}\dag$ & $7.50\pm0.5$ & [8] \\
         XTE\,J1550--564  & -- & -- & -- & -- & $4.38^{+0.58}_{-0.41}$  & [9] \\     
        XTE\,J1650--500  & -- & -- & --  & -- & $2.60\pm0.70$  & [10] \\
        GRO\,J1655--40  &  3.66$\pm$1.01 &  3.74$_{-0.91}^{+ 1.65}$ & 7.41$_{-3.33}^{+ 1.49}$ &  150.6$_{ -12.3}^{+ 14.8}$ & $3.2\pm0.2$ & [11] \\
        MAXI\,J1659--152 & -- & -- & -- & -- &  $8.6\pm3.7$  & [12] \\
        GX\,339--4  & -- &  -- & --  & -- &  $>$5.0  & [13] \\
        H\,1705--250  & --  & -- & -- & -- &  $8.6\pm2.0$  & [14] \\
        XTE\,J1752--223  & -- & -- & -- & -- & $5.75\pm2.25$  & [15] \\
        Swift\,J1753--0127  &  --  &  7.15$_{-2.48}^{+ 3.99}$ &8.42$_{-2.85}^{+ 4.32}$ & 88.5$_{ -39.0}^{+ 97.0}\dag$ & $6\pm2$ &[16] \\
        XTE\,J1817--330 & -- & -- & -- & -- & $5.5\pm4.5$  & [17] \\
        V4641\,Sgr  &  6.62$\pm$1.81  & 6.24$_{-1.30}^{+ 2.04}$ & 8.22$_{-1.18}^{+ 1.03}$ & 64.5$_{ -23.0}^{+105.0}$ & $6.2\pm0.7$ & [18] \\
        MAXI\,J1820+070  & 3.23$\pm$1.14   & 3.46$_{-1.03}^{+ 2.18}$ & 3.82$_{-1.23}^{+ 2.89}$ &80.0$_{ -23.8}^{+  08.6}a$ & -- \\
        XTE J1859+226 & -- & -- & -- & -- & $\gtsim$\,14  & [19] \\
        GRS 1915+105 & -- & -- & -- & 22\,\p\,24$^b$ & $8.6_{-1.6}^{+2.0}$ & [20]  \\
        GS\,2000+251  & -- & -- & -- & -- & $2.7\pm0.7$  & [21] \\
        V404\,Cyg  &  2.28$\pm$0.52 & 2.40$_{-0.53}^{+ 0.95}$ & 2.54$_{-0.64}^{+ 1.32}$ & 51.5$_{ -14.0}^{+ 18.0}$ & $2.39\pm0.14$ & [22] \\
        Cyg\,X--1  &  2.37$\pm$0.18  & 2.38$_{-0.17}^{+ 0.20}$ & 2.39$_{-0.18}^{+ 0.21}$ & 22.8$_{  -03.8}^{+  03.8}$ & $1.86_{-0.11}^{+0.12}$  & [23] \\
 \hline
 \hline

     \end{tabular} 
     ~\par
      {\small
    \item Notes: Col. (2) Parallax-inversion distance. 
      (3) Bayesian distance assuming an exponentially-decreasing space density prior; (4) Assuming a Milky Way space density prior. 
      (5) Peculiar velocity relative to Galactic rotation. (6) Previously published distance from literature, and (7) reference for $r_{\rm lit}$.\\ 
      $^\dag$Assuming $r_{\rm lit}$ in the inference of \vpec. $^a$assuming $\gamma$\,=\,0\,km\,s$^{-1}$. $^b$\citet{reid14_grs1915}.\\
References for $r_{\rm lit}$: [1] \citet{gelino03}; [2] \citet{cantrell10}; [3] \citet{gelino02}; [4] \citet{gelino06};
[5] \citet{wu16_gs1124}; [6] \citet{matasanchez15}; [7] \citet{casares09}; [8] \citet{jonker04}; [9] \citet{orosz11}; [10] \citet{homan06};
[11] \citet{hjellming95}; [12] \citet{kuulkers13}; [13] \citet{heida17}; [14] \citet{jonker04}; [15] \citet{ratti12}; [16] \citet{cadollebel07}; [17] \citet{sala07}; [18] \citet{macdonald14}; [19] \citet{corralsantana11}; [20] \citet{reid14_grs1915}; [21] \citet{jonker04};
[22] \citet{millerjones09}; [23] \citet{reid11}.
 \label{tab:results}
   }
    
  \end{table*}
 
 \end{center}

  \begin{figure*}
 \centering
     \includegraphics[width=85mm]{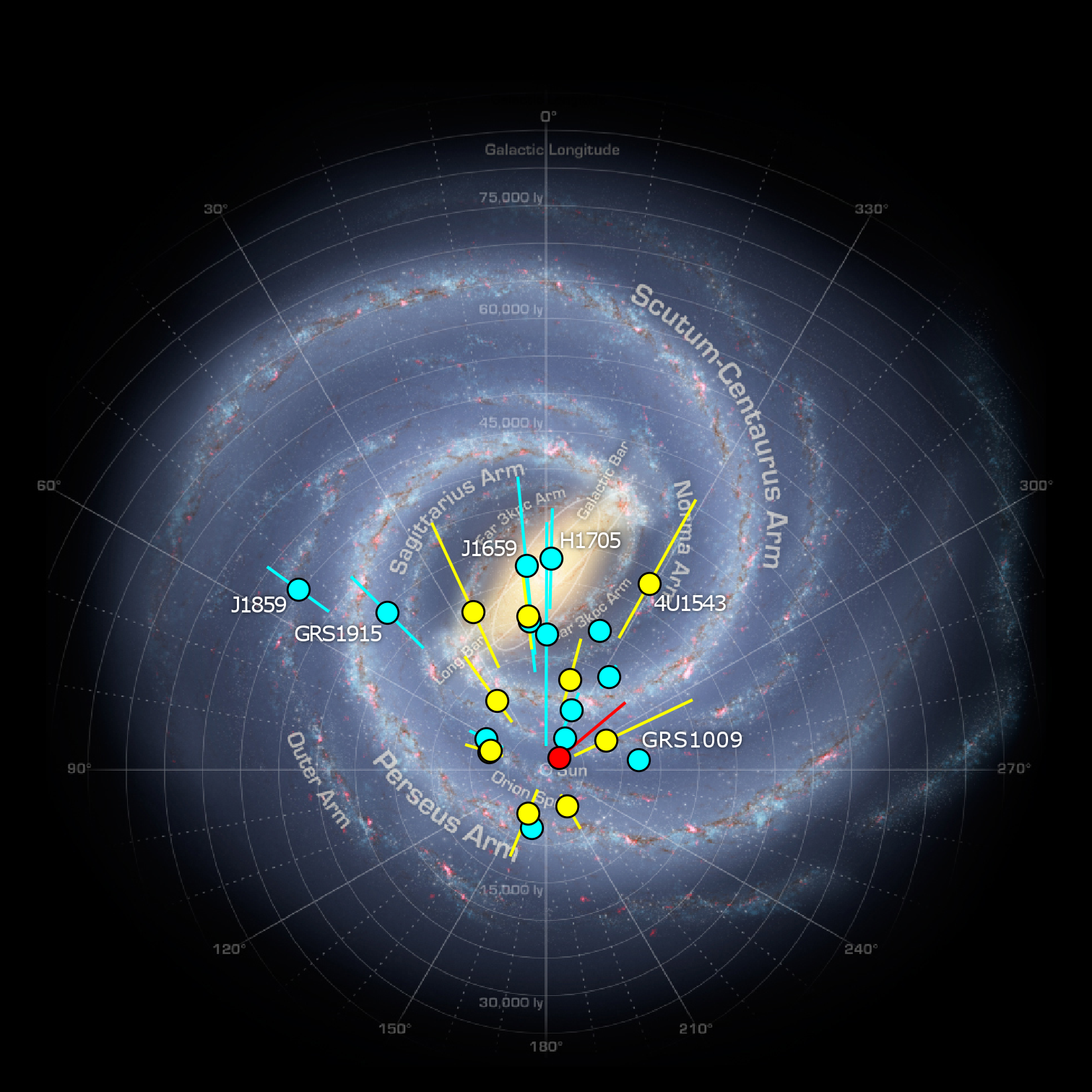} 
     \includegraphics[width=55mm]{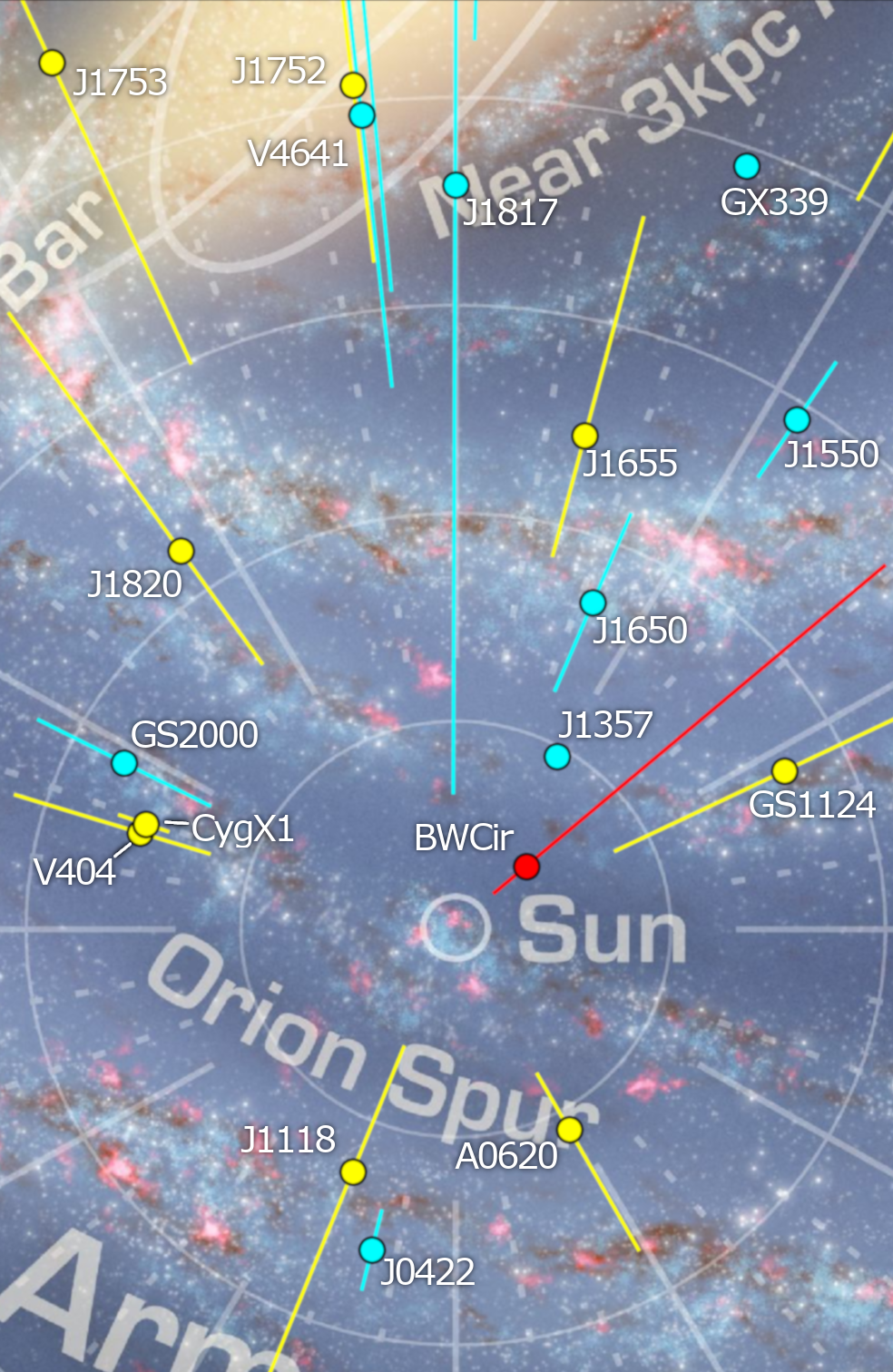}
     \caption{The {\em Gaia} DR2 map of BHXRBs projected on the plane of the Milky Way. Yellow points represent locations calculated using {\em Gaia} data and $r_{\rm exp}$. Cyan points represent objects for which DR2 reports no parallaxes, and their distances are taken from the literature. BW\,Cir is marked in red due to potential caveats on its distance; see Section\,\ref{sec:bwcir} (Milky way image: NASA/JPL-Caltech, ESO, J. Hurt; concentric white circles mark distance increments of 5,000 light-years from the Sun).}
 \label{fig:milkywayplane}
 \end{figure*}

\section{Discussion}
\label{sec:discussion}

\subsection{Distance estimates}
\label{sec:distanceposteriors}

  Table\,\ref{tab:results} shows agreement between the various DR2 distance estimators within measured uncertainties for all systems. The most likely values of $r_{\rm exp}$ are typically straddled by $r_{\rm inv}$ and $r_{\rm MW}$. The high fractional uncertainties on the parallaxes, however, imply that $r_{\rm inv}$ is unlikely to be a reliable distance estimator for the majority of systems. On the other hand, $r_{\rm MW}$ is biased systematically higher than $r_{\rm exp}$ in most cases, with a median excess factor of 1.29. The agreement between $r_{\rm exp}$ and $r_{\rm inv}$ is much closer, with the corresponding factor being 1.05.  
  
  The $r_{\rm MW}$ overestimate likely arises from the fact that the Milky Way prior is constructed by considering the space distribution of X-ray sources, without accounting for effects such as their expected optical flux detectability and the impact of Galactic dust reddening. Consideration of these issues will require modelling the disc and donor star contributions for each source to the {\em Gaia} band, and trying to account for the patchy and uncertain nature of Galactic dust; these refinements are beyond the scope of the present work.

  Given the above comparisons, we adopt $r_{\rm exp}$ as our primary distance estimate for the rest of this work, unless otherwise stated. Such a strategy mimics the adopted choice of an exponential prior in several other works (\citealt{astraatmadja16b}, \citealt{bailerjones2018}), but now with a custom scale length $L$ derived for the dynamically confirmed BHXRB population. 

Regarding comparisons with published distance estimates, the relative uncertainties on the DR2 estimates typically exceed corresponding uncertainties on $r_{\rm lit}$, by a mean factor of at least 3. As already discussed, the ill constrained nature of the geometric parallax measurements is ascribable to the faint nature of the quiescent BHXRB population. Further ongoing astrometric observations should improve these constraints in future data releases. 

We find agreement at a 90\,\% confidence level for 8 of 10 objects between $r_{\rm lit}$ and $r_{\rm exp}$ (Section\,\ref{sec:results}; Fig.\,\ref{fig:distancecomparison}). This is not unexpected, since the scale length $L$ has been derived from the full population of known $r_{\rm lit}$ values. But each source only makes a small contribution to the prior derivation in Section\,\ref{sec:expprior}, so the match for individual sources is encouraging. This is a testament to the DR2 astrometric uncertainties being well characterised on the one hand. Conversely, the DR2 estimates provide a geometric test of the validity of the published distances (based upon a variety of photometric, spectroscopy and astrometric techniques). An additional graphical comparison against $r_{\rm inv}$ and $r_{\rm MW}$ can be found in the Appendix. 

For A\,0620--00 and XTE\,J1118+480, there is only modest disagreement at the 1-$\sigma$ level, and full consistency at the 90\,\% confidence level. 
The astrometric solution for XTE\,J1118+480 is particularly poor, with $\pi/\sigma_\pi$\,$<$\,1, so this should be treated with caution. 
For A\,0620--00, the adopted value of $r_{\rm lit}$ comes from a detailed dynamical model by \citet{cantrell10}, which accounts for the disc contribution to the ellipsoidal orbital light curves, yielding a derived distance of 1.06\,$\pm$\,0.12\,kpc. \citeauthor{cantrell10} show that there is a weak dependence of this distance on the donor spectral type and reddening assumed, with mean values of 0.92--1.26\,kpc possible. The upper end of this range is already consistent with the $r_{\rm exp}$ estimate. We also note agreement of $r_{\rm lit}$ with the $r_{\rm inv}$ estimate, so this mild disagreement could simply be reflective of some intrinsic scatter in the spatial distribution beyond that modelled by our prior assumptions.

The two cases of stronger disagreement are Cyg\,X--1 and BW\,Cir. Cyg\,X--1 is one of the few sources with a prior radio parallax distance, $r_{\rm lit}$\,=\,1.86$_{-0.11}^{+0.12}$\,kpc \citep{reid11}. Our DR2 estimate of $r_{\rm exp}$\,=\,2.38$_{-0.17}^{+0.20}$\,kpc disagrees with this at the 98.8\,\% level assuming a Normal distribution for the radio parallax uncertainties. For BW\,Cir, there is no radio parallax, but the best constraints suggest a distance of $r_{\rm lit}$\,$\gtsim$\,25\,kpc \citep{casares04, casares09}. For this system, we derive $r_{\rm exp}$\,=\,0.70$_{-0.30}^{+3.43}$\,kpc. Only 0.02\,\% of the $r_{\rm exp}$ posterior distribution lies beyond 25\,kpc (see Appendix), implying disagreement between the two estimates of 99.98\,\%. Whereas the DR2 mode estimate is relatively close to $r_{\rm lit}$ for Cyg\,X--1 (differing by a factor of only 1.28), the difference in the case of BW\,Cir is apparently a factor $>$\,30. The causes of these mismatches probably differ between the two cases, and we will explore these further below. 

Fig.\,\ref{fig:milkywayplane} shows our BHXRB sample superposed on an schematic representation of the Galaxy. DR2 $r_{\rm exp}$ estimates are plotted where available, together with $r_{\rm lit}$ values for other systems. This illustration serves to depict the \lq {\em Gaia} view\rq\ of BHXRBs projected on the plane of the Milky Way. 

\subsection{Astrometric fits}
\label{sec:flags}

Table\,\ref{tab:sourceswithastrometry} lists several flags and parameters relevant for assessing the quality of the DR2 astrometric fits. A description of the flags can be found in the reference cited in footnote\,\ref{note:dr2model}. All reported fits are based on 9 or more visibility periods ({\tt visibility\_periods\_used}), i.e. distinct groups of observations separated from other groups by 4 or more days. Small values (e.g. less than 10) could be indicative of additional modelling uncertainties (Ibid.). In only one case (V4641\,Sgr) is the astrometric solution based upon 9 visibility periods, and the DR2 estimated distance in this case agrees with $r_{\rm lit}$. The total number of individual observations spanning all visibility periods exceeds 100 for all sources. So none of these solutions appears to be strongly biased by sparse or irregular sampling.

  The {\tt astrometric\_gof\_al} (hereafter, {\tt gof}) values tabulated represent the \lq Gaussianised $\chi^2$\rq\ goodness-of-fit statistic with zero mean and unit standard deviation (Ibid.). While most are below a value of +3 and thus indicate a reasonable astrometric fit solution, there are a few which are notably above this, e.g. V404\,Cyg ({\tt gof}\,=\,9.5) and Cyg\,X--1 ({\tt gof}\,=\,3.6). 
  So could we explain away the significant offset between $r_{\rm exp}$ and $r_{\rm lit}$ for Cyg\,X--1 based upon a bad astrometric fit? This is not obviously the case, because three other sources with DR2 $r_{\rm exp}$ values matching literature estimates also have an apparently bad astrometric {\tt gof}: XTE\,J1118+480, GRO\,J1655--40 and V404\,Cyg. 
  Furthermore, as we will discuss shortly, the peculiar velocity inferred from the astrometric solution for Cyg\,X--1 matches previous estimates (Section\,\ref{sec:vpec}). 

  Table\,\ref{tab:sourceswithastrometry} does reveal an approximate trend of high {\tt gof} values occurring together with significant {\tt astrometric\_excess\_noise} (hereafter, {\tt noise}), which is a measure of excess residuals in the best-fitting astrometric solution. Such residuals could result from instrumental effects such as image centroiding errors and issues with the telescope attitude, or could be indicative of an intrinsic complexity in the nature of the system under study (Ibid.). In particular, such complexity may arise for our sample of BHXRBs due to \lq orbital wobble\rq, when the flux-weighted centroid of emission wobbles at the orbital period of the binary system.

  An estimate of the maximal possible astrometric wobble may be obtained by assuming that the quiescent flux of BHXRBs is dominated by the companion star. In this case, the angular wobble $w$ is simply $w$\,=\,$2\frac{a_2}{r}$, where $a_2$ is the separation of the companion from the centre of mass and $r$ is the source distance. With knowledge of the system's orbital parameters, this can expressed as

  \begin{equation}
    \label{eq:wobble}
w=2\frac{G M_{1}}{K_2^2 \left( 1+ q \right)^2r} \sin^2 i 
\end{equation}

\noindent
where $M_{1}$ is the mass of the compact object, $K_2$ the velocity amplitude of the companion, $q$ is the mass ratio $\frac{M_2}{M_1}$, and $i$ the binary 
  inclination. Table\,\ref{tab:orbitalseparation} lists the known system parameters together with the expected angular wobble in mas. The distances are taken to be the $r_{\rm lit}$ values, since the aim here is to check the DR2 solutions. For BW\,Cir, $r$ is assumed to be 25\,kpc and $i$\,=\,79\,deg, giving a robust upper limit on $w$. The lower limit on $M_{\rm BH}$ of 7\,\Msun, however, results in a lower limit on $w$. For two systems of those with DR2 astrometric solutions listed in Table\,\ref{tab:sourceswithastrometry}, Swift\,J1753.5--0127 and MAXI\,J1820+070, the orbital solutions are currently not well constrained enough for use here.

  $w$ is found to be larger than the statistical uncertainties on the parallax, $\sigma_\pi$, for both V404\,Cyg and Cyg\,X--1. This indicates that {\em Gaia} may have detected the orbital astrometric wobble in these two systems, and could explain the high {\tt gof} statistic for both sources as well as the significant {\tt noise} for V404\,Cyg. One additional systematic factor for V404\,Cyg could be associated with a bright outburst that the source underwent 2015. This resulted in the disc brightening by several mags in the optical during June and December of that year (e.g. \citealt{kimura16}, \citealt{munozdarias17}). This would introduce brief systematic shifts of the optical light centre from the companion star to the disc, and it will be interesting to search for this effect in the full {\em Gaia} astrometric measurements once those are released.

  Amongst other sources, there could be mild contributions in the case of V4641\,Sgr and GRO\,J1655--40 ($\frac{w}{\sigma_\pi}$\,$\approx$\,0.35 for both). The astrometry of the remaining systems is unlikely to be affected by $w$, even assuming reasonable mass ranges for BW\,Cir.

Cyg\,X--1, however, shows {\tt noise}\,=\,0. This is somewhat surprising given the brightness of this source as compared to V404\,Cyg, but it may be that systematic uncertainties act to suppress the noise. A global systematic error of up to 0.1\,mas in DR2 (e.g. \citealt{luri18}) would easily bring Cyg\,X--1 in line with the radio parallax distance. Based upon the Collaboration recommendation, we have not included this systematic uncertainty in the quoted errors (Ibid.). True offsets between the published radio and the optical DR2 measurements may also arise if the radio jet wobble differs from the optical wobble, as a result of the jet emission arising from an extended, beamed and precessing zone from one side of the jet. The companion star in this case is more massive than the compact object, so the optical wobble is expected to be smaller than any radio one-sided jet wobble. Investigation of this issue will require a detailed comparison of the radio data with the full {\em Gaia} measurements once those are released. Any optical emission from the jet is likely be much weaker than the companion star in this case \citep{rahoui11} and, in any case, optical jet emission should arise on much more compact scales \citep[e.g. ][]{gandhi17}.

For the final discrepant source, BW\,Cir, the {\tt gof} and {\tt noise} parameters are not indicative of a significant issue with the fit. In this case, one further possibility we consider is the impact of correlations between the astrometric fit parameters (e.g. \citealt{lindegren18}, \citealt{luri18}). The DR2 fit solutions include 10 pair-wise correlation coefficients for these; additionally, the DR2 flag {\tt astrometric\_sigma5d\_max} denotes the longest semi-major axis of the 5-d error ellipse when including correlations -- a rough single measure of the effect of the covariances. Amongst our sample of sources with five-parameter fits, BW\,Cir turns out to have the largest value of this flag, by far (Table\,\ref{tab:sourceswithastrometry}), indicative of potential additional covariances. \citet{lindegren18} suggest several checks for accepting or rejecting the DR2 astrometric solutions, including a check on the {\tt sigma5d} parameter -- see their Eq.\,11, where they recommend accepting a solution if {\tt sigma5d\_max}\,$\le$\,(1.2\,mas)\,$\times$\,max[1,\,10$^{0.2(G-18)}$], where $G$ is the average source mag. BW\,Cir passes this check.

In order to test the effect of parameter correlations, we used the Markov Chain Monte Carlo code of \citet{bailerjones_inference} to infer the distance posterior when accounting for covariances between parameters. The code accounts for covariances between parallax and transverse (proper motion) speed by default. We modified this to additionally include covariances on position, incorporating the full DR2 covariance matrix. The prior assumed was our canonical exponentially decreasing space density function for distance, with $L$\,=\,2.17\,kpc. Priors on other parameters were assumed to be uniform. A set of 200 walkers ran over 1500 iterations with an initial burn-in period of 500, using step sizes of 5\,pc in distance, 0.05\,mas\,y$^{-1}$ in proper motion (both RA and Dec), and 0.05\,mas in position (both RA and Dec). Even after accounting for these correlations, we find $r_{\rm mode}$\,=\,1.54$_{-1.25}^{+4.08}$\,kpc, with a distance of 25\,kpc still strongly disfavoured. This posterior distribution is also presented in the Appendix. 

In summary, the astrometric flags do not reveal any obvious spurious solutions at the present time. Cases with apparently bad goodness-of-fit parameters may signal the presence of inherent complexity such as orbital wobble, and confirmation of this will be possible after the final (full) {\em Gaia} data release. The most discrepant case, BW\,Cir, does have have any astrometric flag indicative of a poor solution. This source will be discussed in more detail in Section\,\ref{sec:bwcir}.

\begin{table*}
  \centering
  \caption{Estimated magnitude of orbital wobble\label{tab:orbitalseparation}}
  \begin{tabular}{lcccccr}
    \hline
    \hline
    Source          & $K_2$       &  $i$ & $M_1$ & $q$ & $r_{\rm literature}$& $w$\\
                    & km\,s$^{-1}$&  deg & \Msun &      & kpc                &  mas\\
    (1)             &  (2)        &  (3) &  (4)  & (5)  & (6)                &  (7) \\
    \hline
    A\,0620--00     &  437        &  51  &  6.6\,\p\,0.3 & 0.074 & 1.06            & 0.030 \\
    XTE\,J1118+480  &  709        &  74  &  7.5\,\p\,0.7 & 0.024 & 1.7             & 0.014\\
    GS\,1124--684   &  407        &  43  &  11.0\,\p\,1.75 & 0.079 & 4.95             & 0.010\\
    BW\,Cir         &  279        &$<$\,79$^\dag$ &  $>$7.0$^\dag$ & 0.12 & $\gtsim$\,25$^\dag$              & 0.005 \\
    4U\,1543--475   &  124        &  21  &  9.4\,\p\,1.0 & 0.28 & 7.5             & 0.011\\
    GRO\,J1655--40  &  226        &  69  &  6.0\,\p\,0.4 & 0.42 & 3.2             & 0.028\\
    V4641\,Sgr      &  211        &  72  &  6.4\,\p\,0.6 & 0.66 & 6.2           & 0.014 \\
    V404\,Cyg       &  208        &  67  &  9.0$^{+0.2}_{-0.6}$ & 0.067 & 2.39           & 0.115 \\
    Cyg\,X--1         &   75        &  27  & 14.8\,\p\,1.0 & 1.30 & 1.86 & 0.098 \\
    \hline
    \hline
  \end{tabular}
  ~\par
  Orbital parameters in Cols.\,(2)--(5) are taken from BlackCAT, \citet{casares14}, and papers listed in Tables\,\ref{tab:sample} and \ref{tab:sourceswithastrometry}. Col.\,(6): Literature distances; see Table\,\ref{tab:results}. Col.\,(7) Orbital wobble $w$ computed according to Eq.\,\ref{eq:wobble}. $\dag$Value assumed to be the quoted limit for computing $w$. 
\end{table*}

\subsection{Peculiar Velocities and System Kinetic Energies}
\label{sec:vpec}

DR2 infers \vpec\ values significantly above 100\,km\,s$^{-1}$ for XTE\,J1118+480 and GRO\,J1655--40 (Table\,3). High runaway velocities for both were first inferred by \citet{mirabel01} and \citet{mirabel02} based upon radio and optical ({\em Hubble Space Telescope}) astrometry, respectively. DR2 agrees with this assessment in both cases, though our DR2 estimate is larger than that of \citet{mirabel02} for GRO\,J1655--40 by about 50\% due to apparently higher values of $\mu$. In particular, for $\mu_\delta$, these authors found a value of --4.0\,\p\,0.4\,mas\,y$^{-1}$ whereas DR2 finds --7.44\,\p\,0.09\,mas\,y$^{-1}$. 
  Other published kinematic measurements include V404\,Cyg and Cyg\,X--1. Our DR2 estimate for V404\,Cyg agrees well within uncertainties with the measurement of $\upsilon_{\rm pec}$\,=\,39.9\,$\pm$\,5.5\,km\,s$^{-1}$ by \citet{millerjones09}. Similarly, for Cyg\,X--1, \citet{reid11} estimate $\upsilon_{\rm pec}$\,$\approx$\,21\,km\,s$^{-1}$, in agreement with our DR2 estimate.

First proper motions measurements for 7 other sources are now available from DR2 (Table\,\ref{tab:sourceswithastrometry}). BW\,Cir is identified as a potential high velocity system, with \vpec\ of about $\gtsim$\,100\,km\,s$^{-1}$. At the apparently small most likely distance of $r_{\rm exp}$\,$\ltsim$\,1\,kpc, the peculiar velocity is dominated by the radial velocity of $\gamma$\,=\,103\,km\,s$^{-1}$. Should this distance be an underestimate, the relative contribution of $\gamma$ may drop, but the measured DR2 proper motion of $\mu$\,=\,11.0\,\p\,2.2\,mas\,y$^{-1}$ will then push up \vpec\ (for further analysis on this source, see section\,\ref{sec:bwcir}).

DR2 infers modestly high likely $\upsilon_{\rm pec}$ values of $\sim$\,50--100\,km\,s$^{-1}$ for all other sources: 
A\,0620--00, V4641\,Sgr, MAXI\,J1820+070, GS\,1124--684, 4U\,1543--475 and Swift\,J1753.5--0127. The last two sources have particularly poor parallax constraints ($|\frac{\pi}{\sigma_\pi}|<1$) and we verified that these sources lie in the class of intermediate to modestly high \vpec\ values irrespective of whether we use adopt $r_{\rm lit}$ or $r_{\rm exp}$ as their distance estimator. 

\begin{figure*}
   \centering
   \includegraphics[width=8.75cm]{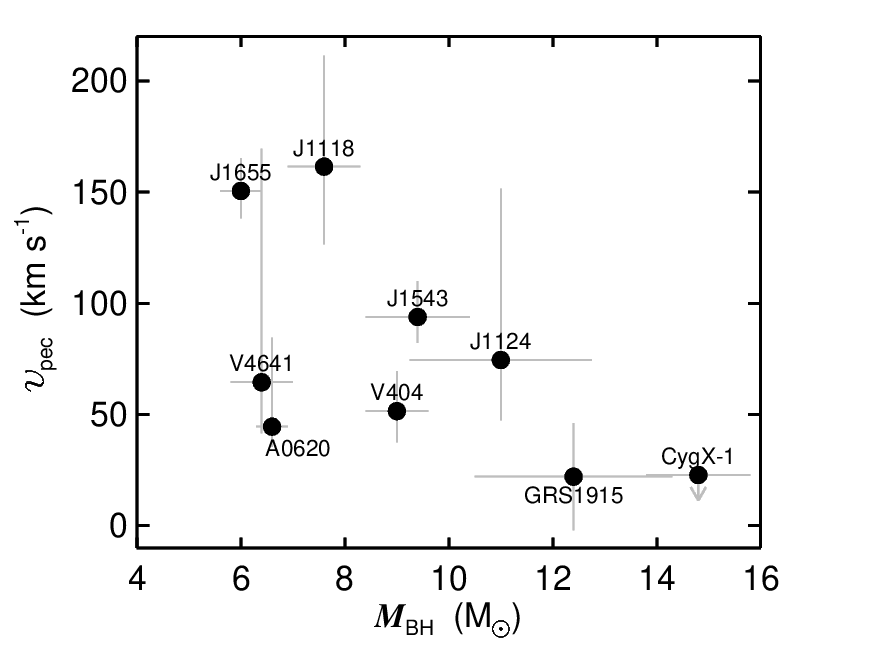}
   \hfill
   \includegraphics[width=8.75cm]{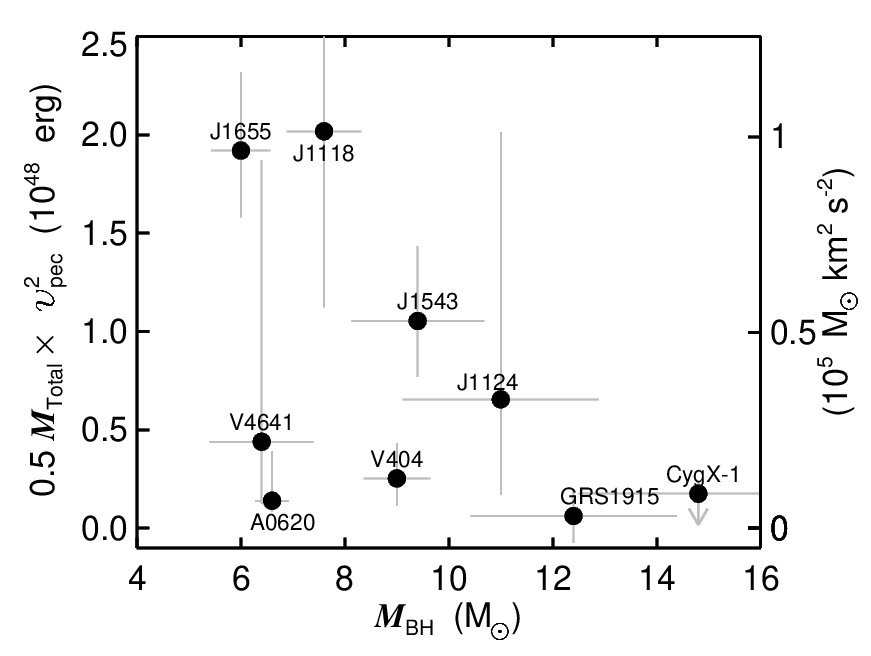} 
   \caption{Black hole mass (\mbh) vs. peculiar velocity ({\em left}), and vs. system kinetic energy relative to that expected if the source were undergoing pure Galactic rotation ($K_{\rm pec}$; {\em right}).}
   \label{fig:momentum}
\end{figure*}

In Fig.\,\ref{fig:momentum}, we present correlations of \mbh\ against \vpec, and against system kinetic energy, $K_{\rm pec}$\,=\,$\frac{1}{2}M_{\rm Total}$\,$\times$\,\vpec$^2$ where $M_{\rm Total}=M_1+M_2$. Systems with a lack of secure black hole masses (BW\,Cir and MAXI\,J1820+070) were not included. Furthermore, as pointed out by \citet{mirabelrodrigues03}, the peculiar velocity of Cyg\,X--1 relative to its parent association of massive stars, Cyg\,OB3, is likely even smaller than the velocity relative to the Galaxy. They quote a value lower than 9\,\p\,2\,km\,s$^{-1}$. Therefore, we treat our value of \vpec\ for Cyg\,X--1 as an upper limit here. The median \vpec\ of this sample is 65\,km\,s$^{-1}$ and the median $K_{\rm pec}$ is 5\,$\times$\,10$^{47}$\,erg. Assuming typical supernova energies of $E_{\rm SN}$\,$\sim$\,10$^{51}$\,erg \citep{kasenwoosley09}, $K_{\rm pec}$ represents 0.05\,\% of $E_{\rm SN}$.

The figure reveals potential weak anti-correlations in both cases: Spearman's rank correlation coefficient $r_{\rm S}$\,=\,--0.53 and --0.43 for the relations against \vpec\ and against $K_{\rm pec}$, respectively. But with insignificant corresponding $p$-values of only 0.14 and 0.24, respectively, it is likely that anti-correlation trends as steep as observed arise simply as a result of chance, within this small sample. 

One can ask the alternative question of how likely is it that the correlation coefficients $r_{\rm S}$ (or equivalently, the slopes) are {\em negative} (i.e. consistent with an {\em anti}-correlation), irrespective of their exact values. In order to answer this, we carry out resampling of the \vpec, \mbh\ and $K_{\rm pec}$ values to generate 10,000 randomised ensembles each comprising 9 objects, by using the known system parameter uncertainties together with the \vpec\ posterior distributions for the resampling. For the case of Cyg\,X--1, our measured \vpec\ was treated as an upper limit (as mentioned above) for uniform sampling of smaller randomised values. Each random ensemble was analysed as before, resulting in a distribution of 10,000 corresponding randomised $r_{\rm S}$ values for both correlations. The strength of any apparent anti-correlation can then be quantified by checking the fraction of $r_{\rm S}$ values that are negative. With this test, we find that the correlation coefficients are negative for 99.9\,\% and 98.8\,\% of the simulated trends against \vpec\ and against $K_{\rm pec}$, respectively. These percentages are consistent with more statistically significant anti-correlations than implied by the raw sample $p$-values in the preceding paragraph. This is because several of the individual source \vpec\ uncertainties in the low-to-intermediate \mbh\ regime in Fig.~\ref{fig:momentum} are skewed towards high positive \vpec\ values which, in turn, would favour any trends being skewed towards negative (anti)correlation coefficients.

We also used these 10,000 randomised ensembles to fit linear regression slopes to the trends together with uncertainties. For the regression trend of \vpec\ against \mbh, we find a slope of --10.6$_{-6.2}^{+4.6}$\,km\,s$^{-1}$\,\Msun$^{-1}$, whereas for the regression of $K_{\rm pec}$ against \mbh, we find a slope of --0.09$_{-0.14}^{+0.10}$\,$\times$\,10$^{48}$\,erg\,\Msun$^{-1}$. The quoted values are the modes of the ensemble slope distributions, and the uncertainties refer to the corresponding 68.3\,\%\ highest density intervals.

  Similar correlations have been investigated before using smaller subsets of objects \citep{millerjones14, mirabel17review}, and have the potential to constrain black hole (BH) formation scenarios. Theoretical models suggest potentially distinct formation channels of direct collapse for high mass BHs, and delayed supernova fallback onto central neutron stars for low mass BHs (e.g., \citealt{fryerkalogera01}). In such cases, inverse correlations between the imparted natal kicks and system mass may be expected. Our result of putative anti-correlations is intriguing and is consistent with such formation channels. We also note that our median \vpec\ value of $\approx$\,80\,km\,s$^{-1}$ would easily exceed the kick velocities inferred for BH formation. But we caution against drawing substantive conclusions at present. Despite the exquisite DR2 constraints on $\mu$, the relatively small sample size and the uncertainties on $\pi$ and \mbh\ still contribute significant scatter. In particular, only the slope for the former correlation (\vpec\ against \mbh) are significantly negative using the simulations above; any anti-correlation of $K_{\rm pec}$ against \mbh\ remains insignificant in the present data. This could simply be because the definition of $K_{\rm pec}$ conflates uncertainties of system mass together with peculiar velocity. In any case, follow up of these sources to further mitigate \mbh\ as well as $K_{\rm pec}$ uncertainties, as well as enlarging the base sample, will be needed to confirm or rule out the validity of these trends.


\subsection{The curious case of BW\,Cir}
\label{sec:bwcir}

BW\,Cir, or GS\,1354--64, is a dynamically confirmed black hole \citep{casares09}. Previous estimates on its distance have placed it at a possible distance of $\gtsim$\,25 kpc \citep{casares04,casares09}, on the other side of the galaxy. There are good reasons for such an inference, as discussed in the above papers, including the donor star classification and expected high luminosity given its relatively long orbital period (2.5\,days), as well as the fact that its radial velocity matches with the Galactic differential speed at that distance. However, {\em Gaia} apparently disagrees with this estimate strongly, placing it at a likely distance of $r_{\rm exp}=0.7_{-0.3}^{+3.4}$\,kpc (see discussion in Section\,\ref{sec:distanceposteriors}). For this source, the Milky Way prior based estimate $r_{\rm MW}=7.0_{-6.3}^{+3.4}$\,kpc. Although apparently less discrepant with respect to $r_{\rm lit}$, note that the lower confidence interval for $r_{\rm MW}$ shows a significant tail towards small distances, and at the upper end, values of more than 25\,kpc still only have a probability of being favoured at $\approx$\,0.5\,\% (we refer to the Appendix for a plot of the distance posterior distribution). As discussed in Section\,\ref{sec:flags}, there is no reason to suspect a spurious solution based upon the astrometric solution flags, so this discrepancy is puzzling.

\begin{figure}
 \centering
     \framebox{\includegraphics[width=0.45\textwidth]{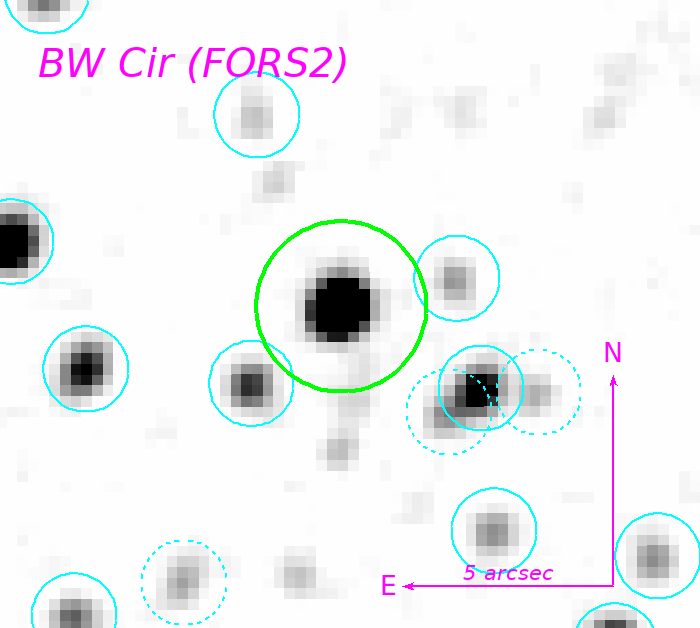}}
     \caption{FORS2/VLT $V$-band image of the field of BW\,Cir. The source counterpart is within the green 2\,\arcsec\ radius circle. Neighbouring sources are highlighted within cyan apertures of radius 1.\arcsec\ Sources with a reported parallax in DR2 are outlined by the continuous circles; dashed circles denote sources without a reported parallax. The image was taken on 2015 Jun 26 near the beginning of a source outburst.}
 \label{bwcir}
 \end{figure}

In order to rule out any systematics related to source identification, we present in Fig.\,\ref{bwcir} the DR2 counterparts found within a larger search radius around the source, overlaid on a ground-based (VLT/FORS2) optical image of the field. 
There is little doubt that {\em Gaia} correctly identifies BW\,Cir and its nearest neighbours, implying no immediate identification issues. 

\begin{figure}
   \centering
   \includegraphics[width=8.5cm]{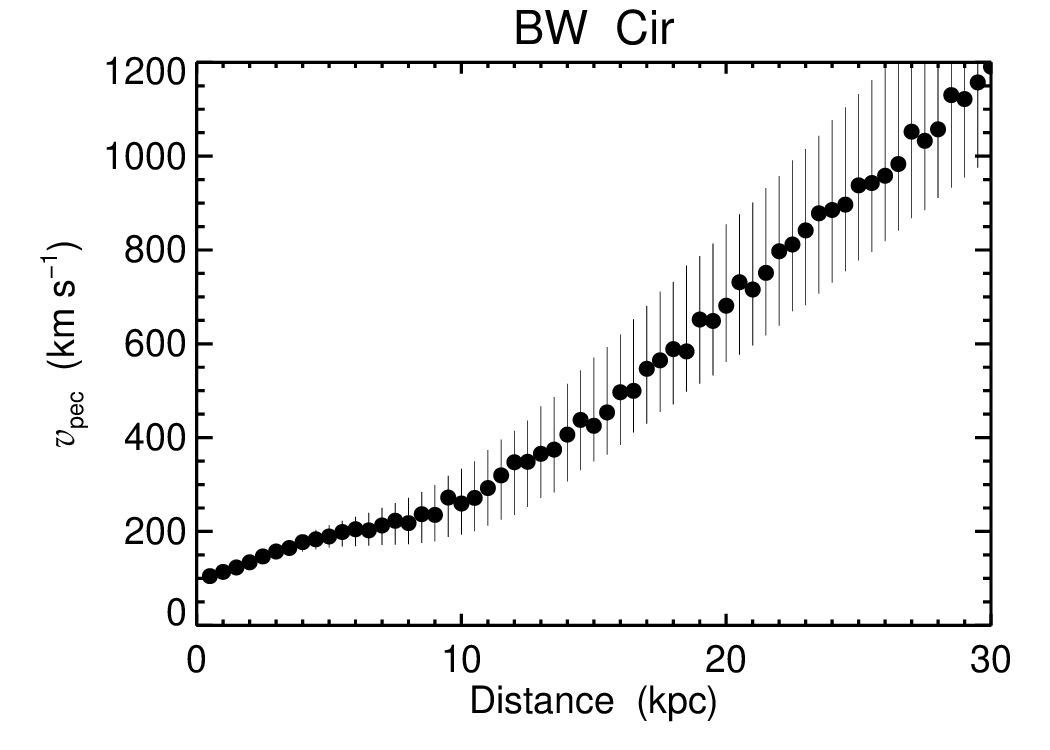} 
   \caption{Expected \vpec\ for BW\,Cir using the DR proper motions and known radial velocity, for a range of possible distances over 0.5--30\,kpc.}
   \label{fig:bwcir_vpec_dist}
\end{figure}

At its inferred DR2 distance, \vpec\ is dominated by the systemic heliocentric radial velocity $\gamma$. However, we note that if the source identification were correct but the distance were significantly larger, the DR2 proper motions would start pushing up \vpec\ to several hundred km\,s$^{-1}$ or more. This trend is displayed in Fig.\,\ref{fig:bwcir_vpec_dist}. While this cannot be ruled out a priori, it adds additional support to a closer distance. We also note that a high proper motion was measured in the Hot Stuff for One Year Catalogue \citep{altmann17}, although that is now superseded in terms of precision by DR2. 

The one additional potential systematic that is difficult to quantify at present is the effect of flux variability. BW\,Cir underwent an outburst for several months in 2015, during which time it brightened to $V$\,$\approx$\,17.6\,mag (e.g. \citealt{koljonen16}). Amongst the full DR2 data release, there are a very small number of cases where the pipeline photometric processing fails in the presence of significant intrinsic variability -- one prominent case highlighted by the {\em Gaia} team is that of RR\,Lyrae, whose $G$-mag as well as parallax are reported to be incorrect \citep{gaiadr2}. For this source, however, both the {\tt gof}\,=\,212.5 and the {\tt noise}\,=\,3.60 flags strongly indicate a bad fit, unlike BW\,Cir. 
  Moreover, the reported $G$-mag of BW\,Cir is not dramatically different from that expected in quiescence \citep{blackcat}. So any adverse impact of outburst flux variability on the DR2 astrometric solution remains unclear at this stage.

In summary, the DR2 identification for BW\,Cir appears reliable, and the parallax and proper motion argue for this source being a nearby system. Systematic uncertainties on the parallaxes and those related to variable source processing would not obviously result in spurious astrometry. 
A nearby distance would also bring the source quiescent X-ray luminosity $L_{\rm X}$ in line with the $L_{\rm X}$--$P_{\rm orb}$ (orbital period) trend for black hole binaries (cf. \citealt{reynoldsmiller11}, who point out that the source is overluminous by more than an order of magnitude in X-rays as compared to other BHXRBs, if at a distance of 25\,kpc), and also revives the possibilities of the source being associated with Cen\,X--2 (cf. \citealt{casares04, casares09}). 

But we caution that further investigation is warranted in order to (1) carry out other complementary checks on the distance -- e.g. with future radio facilities -- and, (2) to understand implications for the quiescent (and outburst) source nature. In particular, taking the distance at the low end of the distribution has major implications for the amount of light that can be produced by the donor star, and, in turn the radius of the donor star.  The donor star's period is well established \citep{casares04, casares09} and its spectral temperature classification is unlikely to be subject to significant revision, so changing the donor star distance changes its radius.  If the donor star is then also still assumed to fill its Roche lobe, then the mass must be $\sim25^3$ times smaller for a distance of 1 kpc than for a distance of 25 kpc in order to maintain the same density, and the same optical flux for the donor star.  This would yield a donor mass of $\sim5\times10^{-5}$\,\Msun, unreasonably small.  While the donor may potentially be highly bloated, and overluminous for its mass, it is unlikely to be much less than $\sim\,0.1$\,\Msun.  For such a star to be filling its Roche lobe, the distance could be changed by only a factor of about 2 from the 25 kpc approximate estimate of \citet{casares09}.  Some additional flexibility can be derived by allowing the fraction of the optical light coming from the donor star to be a bit smaller than the 30--50\,\% range for different quiescent fluxes assumed by \citet{casares09}, but given the detectability of ellipsoidal modulations from the source in the presence of other variability, and the detectability of the stellar absorption lines, the veiling of the donor star by accretion light cannot be too extreme. Similarly, the system could potentially be wind-fed, and underfilling its Roche lobe, but it could only underfill the Roche lobe by a small amount while still having a tidally locked donor star that produces ellipsoidal modulations. 

The hypothesis of a large distance creates a different problem:
a space velocity that is potentially above the escape velocity from
the Galaxy despite the source's location quite close to the Galactic
Plane. The posterior distributions do allow distances of order $\sim$\,1--10\,kpc with a reasonable probability and a large, but not excessive, space velocity. It could be that the source just happens to lie at such a distance, in the tail of the likelihood distribution $P(\pi|r)$. This would partially reconcile the various distance, stellar classification, and space velocity constraints. But such a solution would remain an unsatisfactory compromise until further data can be obtained, and is speculative at this stage. 
Further data releases from {\em Gaia} should help with improving both the statistical uncertainties, as well as understanding the potential systematics discussed above. We also suggest that other high angular resolution measurements (e.g. with future radio facilities, or pushing current sub-mm facilities to deep limits) to corroborate or revise the DR2 parallax and motion should be carried out.

\subsection{\texorpdfstring{MAXI\,J1820+070}{MAXI J1820+070}}
\label{sec:maxij1820}
MAXI\,J1820+070 is a recently-discovered transient detected with MAXI on March 11, 2018 \citep{kawamuro18} following an optical brightening reported by the ASASSN survey \citep[][ where the designation is ASASSN-18ey]{denisenko18}. The system peak brightness lay amongst the brightest sources in the extra-Solar X-ray sky, and it was observed by many observatories across the electromagnetic spectrum. The system is believed to host a black hole based upon a variety of spectral and timing characteristics, and displayed an exceptionally bright outburst with plenty of multiwavelength variability in the hard X-ray state (e.g., \citealt{baglio18_atel, uttley18, g18_atel}). Its optical position is reported to be consistent with a star which appeared in {\em Gaia} DR1 with nominal astrometric uncertainty of a few mas, but its distance had not previously been estimated.  

The adopted DR2 distance estimate of $r_{\rm exp}$\,=\,3.46$_{-1.03}^{+2.18}$\,kpc corresponds to a height of $z$\,=\,0.61$_{-0.18}^{+0.38}$\,kpc above the Galactic plane, placing the source within the thin disc population of XRBs, though the posterior probability tail of allowed distances could place the source in the thick disc ($z$\,$>$\,1\,kpc).

  The bolometric (1--100\,keV) X-ray flux during this outburst peaked at $F_{1-100}$\,$\approx$\,1.8\,$\times$\,10$^{-7}$\,erg\,s$^{-1}$\,cm$^{-2}$ in March 2018 \citep{shidatsu18}. At the DR2 distance, this yields a luminosity $L_{1-100}$\,$\approx$\,2.6$_{-1.3}^{+4.3}$\,$\times$\,10$^{38}$\,erg\,s$^{-1}$. Assuming a black hole mass \mbh\,=\,10\,\Msun\ implies a peak hard state luminosity of $\approx$\,20$_{-10}^{+34}$\,\% of the Eddington luminosity. Models of hot accretion flows suggest that the hard-to-soft state transition in rising XRB outbursts will occur at Eddington ratios of $\sim$\,1--10\,\%, though observationally, the range is seen to be broader than this (e.g. \citealt{done07}). During the present outburst, the soft state transition occurred after the source had declined from peak, in July 2018 \citep{homan18_atel_j1820_soft}. This places MAXI\,J1820+070 amongst systems capable of maintaining high Eddington ratio hard states. The present uncertainty on the DR2 distance, however, does not allow us to rule out a lower range of Eddington ratios. This source is amongst the brighter of the BHXRBs detectable by {\em Gaia} in quiescence, so future {\em Gaia} releases are expected to tighten the constraints on distance and improve our understanding of the outburst evolution.

MAXI\,J1820+070 illustrates the real power of {\em Gaia}, in that geometric distance estimates are now possible for newly discovered transients (with optical counterparts brighter than $G$\,$\approx$\,20.5 or so), giving an incredibly useful first distance estimate in the absence of detailed spectroscopic follow-up. 

\begin{figure}
 \centering
     \includegraphics[width=85mm,angle=0]{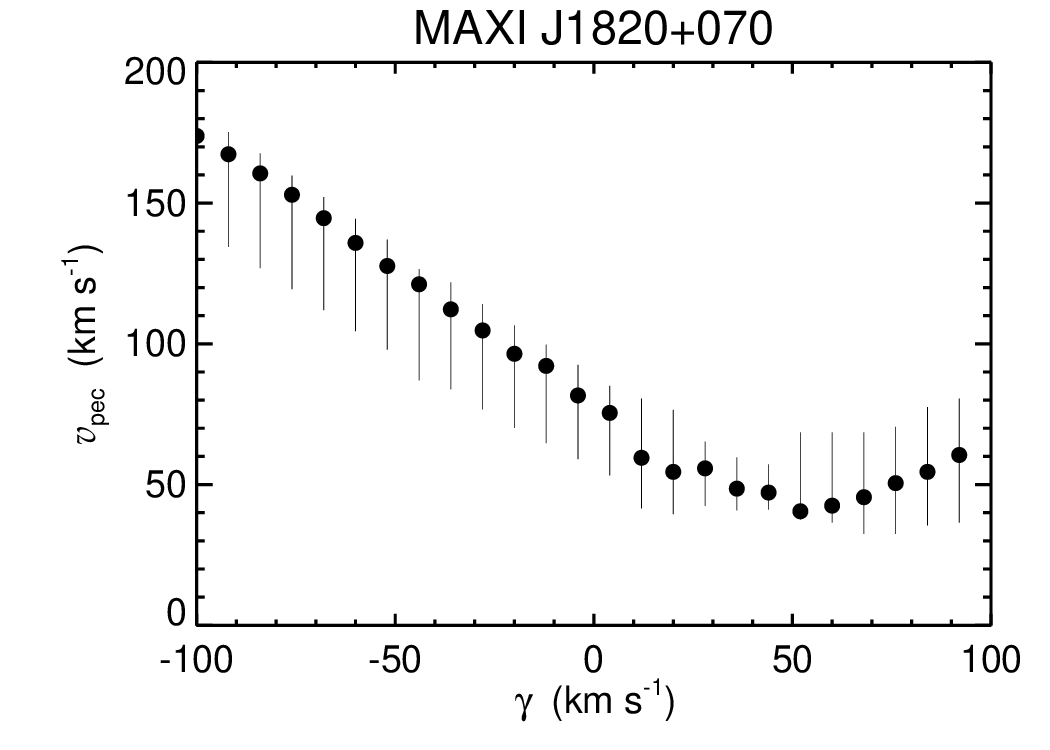}
     \caption{Variation of \vpec\ for MAXI\,J1820+070, for an assumed range of (currently unknown) radial velocities $\gamma$\,=--100 to +100\,km\,s$^{-1}$.
       \label{fig:maxij1820_vpec_distance}}
 \end{figure}

Finally, we note that its present (projected) velocity is mostly parallel to the Galactic plane (see Fig\,\ref{fig:hammer}), also suggesting that the source does not venture significantly beyond the disc. DR2 also delivers a $\approx$\,30-$\sigma$ precision on the proper motion of the source. Despite the absence of a published systemic radial velocity ($\gamma$) at present, the proper motion already suggests that the source must have at least a mildly high peculiar velocity at the inferred likely distance. The value of \vpec\,$\approx$\,80\,km\,s$^{-1}$ for this source (Table\,3) is based upon assuming that $\gamma$\,=\,0\,km\,s$^{-1}$ (cf. other sources in Table\,2). Varying $\gamma$ over the range of --100 to +100\,km\,s$^{-1}$ changes \vpec\ between $\approx$\,50--140\,km\,s$^{-1}$. This trend is shown in Fig.\,\ref{fig:maxij1820_vpec_distance}.

\section{Summary}

We have investigated {\em Gaia} DR2 properties for a sample of Galactic X-ray transients that are either dynamically confirmed or likely accreting black hole X-ray binaries. A summary of our findings is as follows:

\begin{enumerate}
\item DR2 detects the optical counterparts to 18 systems of a sample of 24 BHXRBs. The median mag of the detected sample is $G$\,=\,19.35\,mag, i.e. towards the fainter end of the brightness range covered by DR2 (Section\,\ref{sec:results}). 
\item Almost all secure DR2 counterparts are within 1\,\arcsec\ of published source positions. Sources at greater separation are likely to be unassociated neighbours, and we highlight one case of GRS\,1009--45 where a widely used public database has made a spurious association.
\item Using best literature distance estimates extracted from a variety of techniques, we derive a characteristic expected exponential scale length of this BHXRB population of $L$\,=\,2.17\,\p\,0.12\,kpc. This is smaller than the scale length of XRBs in the Galaxy based upon X-ray detections (Section\,\ref{sec:priors}), likely a result of the selection effects that go into defining this well-characterised population in the optical. One should keep in mind these systematic selection effects when considering the appropriateness of this scale length for other BHXRB studies. 
\item The DR2 geometric distance estimates are in fair agreement with literature estimates. Of 10 objects with literature estimates, there is agreement at the 90\,\% confidence level for 8, despite significant parallax measurement uncertainty in most cases (Fig.\,\ref{fig:distancecomparison}, Section\,\ref{sec:distanceposteriors}). This is encouraging, and provides a first geometric test of many literature estimates that are derived from a variety of photometric and spectroscopy techniques.
\item The two cases of strong disagreement are Cyg\,X--1 and BW\,Cir. The former mismatch (Cyg\,X--1) may be attributable either to intrinsic orbital wobble, or to systematic pipeline measurement uncertainties (Section\,\ref{sec:flags}). There is no immediate cause for the discrepancy of BW\,Cir, and we discuss this case at length (Section\,\ref{sec:bwcir}) ruling out any obvious source identification issue, pipeline related artefact, and astrometric fit covariances. Other potential issues that we are unable to rule out at present include outburst flux variability, and chance occurrence. Future {\em Gaia} releases will be crucial for verifying or rejecting the present DR2 solution, and we also highlight the need for complementary multiwavelength follow up in the mean time.
\item DR2 presents first proper motions measurements for 7 sources, in addition to 4 sources for which literature estimates exist, and these are used to study the kinematics of this population. Proper motions are typically measured much more precisely than parallaxes (Section\,\ref{sec:results}).
\item Combining proper motions with radial velocities, the three-dimensional DR2 peculiar velocities (relative to Galactic rotation) of BHXRBs are estimated, and found to exceed \vpec\,$\approx$\,50\,\,km\,s$^{-1}$ for 9\,sources (Section\,\ref{sec:vpec}). The median kinetic energy of peculiar motion is $K_{\rm pec}$\,$\approx$\,5\,$\times$\,10$^{47}$\,erg, or about 0.05\,\% of typical supernova explosion energies. BW\,Cir is identified as a potential high velocity system, with \vpec\ of about $\gtsim$\,100\,km\,s$^{-1}$ (Section\,\ref{sec:bwcir}). 
\item Tests for dependence between \mbh\ and \vpec\ shows a potential anti-correlation, as may be expected in mass-dependent black hole kick formation channels (Section\,\ref{sec:vpec}). However, we do not find a significant dependence of $K_{\rm pec}$ against \mbh\ in the present sample, and caution that larger samples will be needed to place the present weak trends on a more robust footing.
  \end{enumerate}

\noindent
Since Galactic X-ray transients typically lie in the faint regime during optical quiescence, the DR2 distance estimates have larger uncertainties as compared to dedicated photometric and spectroscopic studies of individual objects. However, they provide very useful, independent geometric estimates that can be used to validate previous studies, and these estimates will improve with future data releases. But the real advantage of the mission will be in providing immediate distance estimates for new outbursting binaries in years to come, as demonstrated in the case of the recent transient MAXI\,J1820+070 (Section\,\ref{sec:maxij1820}).

\section{Acknowledgements}

This work has made use of data from the European Space Agency (ESA) mission {\it Gaia} (\url{https://www.cosmos.esa.int/gaia}), processed by the {\it Gaia} Data Processing and Analysis Consortium (DPAC, \url{https://www.cosmos.esa.int/web/gaia/dpac/consortium}). Funding for the DPAC has been provided by national institutions, in particular the institutions participating in the {\it Gaia} Multilateral Agreement. AR acknowledges a Commonwealth Rutherford Fellowship, JP is part supported by funding from a University of Southampton Central VC Scholarship. PG acknowledges support from STFC (ST/R000506/1). This project benefitted from funding by a UGC-UKIERI Phase 3 Thematic Partnership award. We thank J. Casares and P.A. Charles for discussions on the nature of BW\,Cir, and all coauthors of the BlackCAT catalogue for a useful public resource, in particular J. Corral-Santana. PG also thanks J.C.A. Miller-Jones, R. Plotkin, M. Reynolds, A. Bahramian, J. Steiner, J. Orosz, G. Rate, M. Reid, A.W. Shaw and S. Hodgkin for additional discussions following initial upload of the manuscript to the arXiv server. This research has made use of observatory archival image, including {\em HST}/MAST, ESO, 2MASS, PanSTARRS1 and NASA/IRSA. The authors acknowledge thorough and useful comments from an anonymous reviewer that helped to improve and broaden the scope of the discussion herein. Finally, we acknowledge prior discussions with Jeffrey McClintock on the nature of Cen\,X--2 and its potential association with BW\,Cir.

\bibliographystyle{mnras}
\bibliography{gandhi18b_rev3}
\label{lastpage}

\setcounter{figure}{0}
\makeatletter 
\renewcommand{\thefigure}{A\@arabic\c@figure}
\makeatother
\setcounter{equation}{0}
\makeatletter 
\renewcommand{\theequation}{A\@arabic\c@equation}
\makeatother
\setcounter{table}{0}
\makeatletter 
\renewcommand{\thetable}{A\@arabic\c@table}
\makeatother
\setcounter{section}{0}
\makeatletter 
\renewcommand{\thesection}{A}
\setcounter{subsection}{0}
\makeatletter 
\renewcommand{\thesubsection}{A\@arabic\c@subsection}
\makeatother

\newpage
\section{Appendix}

\subsection{Finding charts for sources with multiple counterparts}

\begin{figure*}
 \centering
     \framebox{\includegraphics[width=85mm,angle=0]{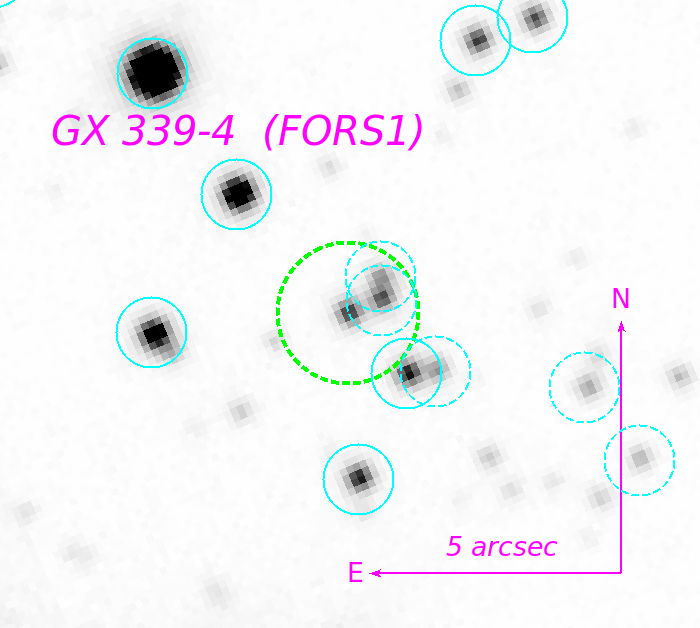}}
     \framebox{\includegraphics[width=85mm,angle=0]{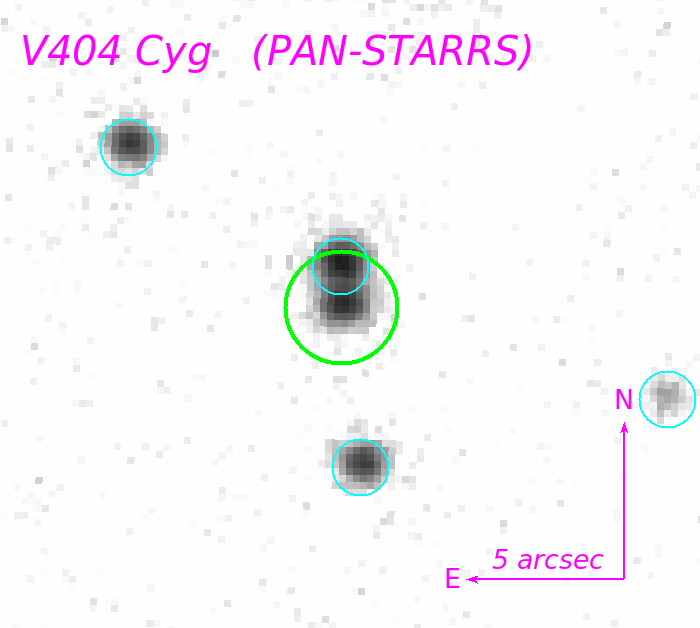}}
     \caption{Finding charts for the fields of GX\,339--4 ({\em Left}) and V404\,Cyg ({\em Right}), showing the associated DR2 counterpart at the centre of the thick continuous green circle (2\,\arcsec\ radius), and neighbouring sources catalogued in DR2 outlined by the smaller cyan circles. The GX\,339--4 image is an archival dataset from VLT/FORS1 in the $r$(Gunn) filter taken on 2000 June 04, while the V404\,Cyg image is from PAN-STARRS in the $g$ filter taken on 2012 February 09. Sources with a reported parallax in DR2 are outlined by the continuous circles; dotted circles denote sources without a reported parallax.
       \label{fig:gx339}}
 \end{figure*}

\begin{figure}
 \centering
     \framebox{\includegraphics[width=85mm,angle=0]{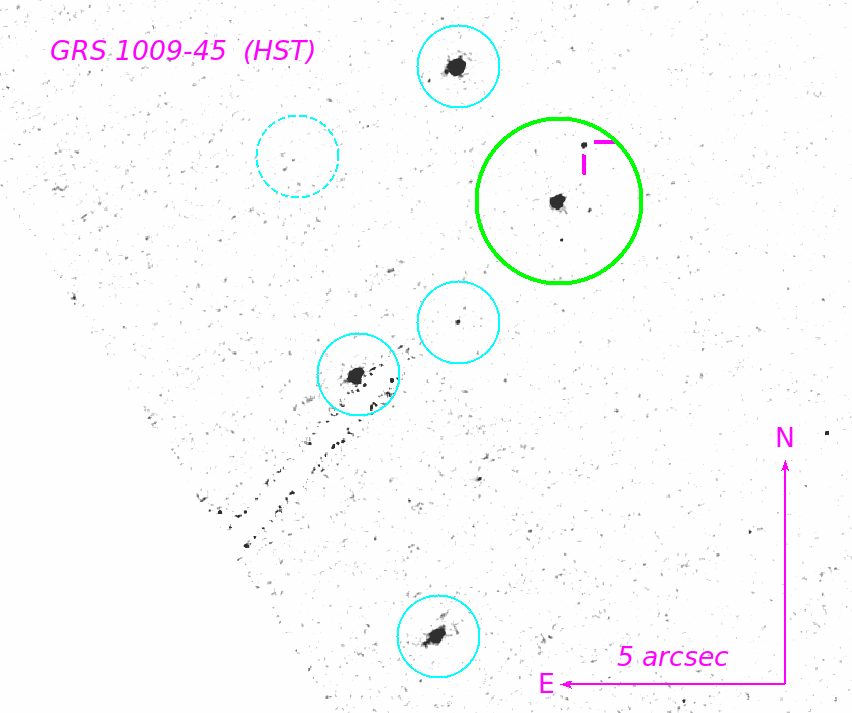}}
     \caption{Finding chart for GRS\,1009--45 from the {\em Hubble Space Telescope}, showing the correct counterpart in the magenta cross-hair and the nearby confused counterpart in green. Other neighbouring sources from DR2 are highlighted in cyan. Sources with a reported parallax in DR2 are outlined by the continuous circles; dotted circles denote source without a reported parallax.
       \label{fig:grs1009}}
 \end{figure}

Finding charts of the fields of sources with multiple potential counterparts are presented in Figs.\,\ref{fig:gx339} and \ref{fig:grs1009}. The correct counterpart is identified in the large central green circle. In the latter figure, the correct counterpart to GRS\,1009--45 is not detected in DR2. 

\subsection{Milky Way Prior}

Eqs.\,(4), (5) and (6) from \citet{grimm02} are reproduced here for clarity, with differences explained below. 

\begin{equation}
  \rho_{b} = \rho_{b}^0 \cdot \left(\frac{\sqrt{(r_{\rm Gal}^{||})^{2} +
                 \frac{z^{2}}{q^{2}}}}{r_0}\right)^{-\gamma} \cdot
                 {\rm exp}\left(-\frac{(r_{\rm Gal}^{||})^{2} +\frac{z^{2}}{q^{2}}}{r_t^2}\right)
\label{eq:bulge}
\end{equation}
\begin{equation}
  \rho_{d} = \rho_{d}^0 \cdot
  \exp\left(-\frac{r_m}{r_d}-\frac{r_{\rm Gal}^{||}}{r_d} - \frac{|z|}{r_z}\right)
\label{eq:disc}
\end{equation}
\begin{equation}
  \rho_{h} = \rho_{h}^0 \cdot \frac{\exp(-b_h \cdot
  (\frac{r_{\rm Gal}}{R_{e}})^{1/4})}{\left(\frac{r_{\rm Gal}}{R_{e}}\right)^{7/8}},
\label{eq:sphere}
\end{equation}
where $\rho_{b}$, $\rho_{d}$ and $\rho_{h}$ represent densities in the bulge, disc, and halo, respectively. $r_{\rm Gal}^{||}$ is the distance in the plane from the Galactic
centre. $z$ is the height above the plane (cylindrical coordinates), and $r_{\rm Gal}$ is the direct distance from the Galactic centre (spherical coordinates). The constants are listed in Table\,\ref{tab:constants}. The normalisations are chosen so that the density of the disc at the position of the Sun satisfies $\rho_{h} (r_{\rm Gal}^{||}=R_{\odot}, z=z_{\odot})$\,=\,$\frac{1}{500}$\,$\cdot$\,$\rho_{d} (r_{\rm Gal}^{||}=R_{\odot}, z=z_{\odot})$, and the total mass of the bulge is 1.3\,$\times$\,10$^{10}$\,\Msun. The integrated disc:bulge:halo ratios are then scaled to be 2.0:1.0:0.8. These reproduce the assumptions adopted in \citet{grimm02}. 
All distances are in kpc. 

There is, however, one noteworthy difference with respect to \citet{grimm02}. The first exponential term in Eq.\,\ref{eq:disc} above represents the contribution of the interstellar medium to the disc. While \citet{grimm02} list this as decreasing with $r_{\rm Gal}^{||}$, this does not match the final published version of the model in \citet{dehnenbinney98}, and appears to be a typographical error. We use the \citet{dehnenbinney98} prescription with the constant $r_m$ instead. 

\begin{figure*}
   \centering
     \includegraphics[width=8.5cm]{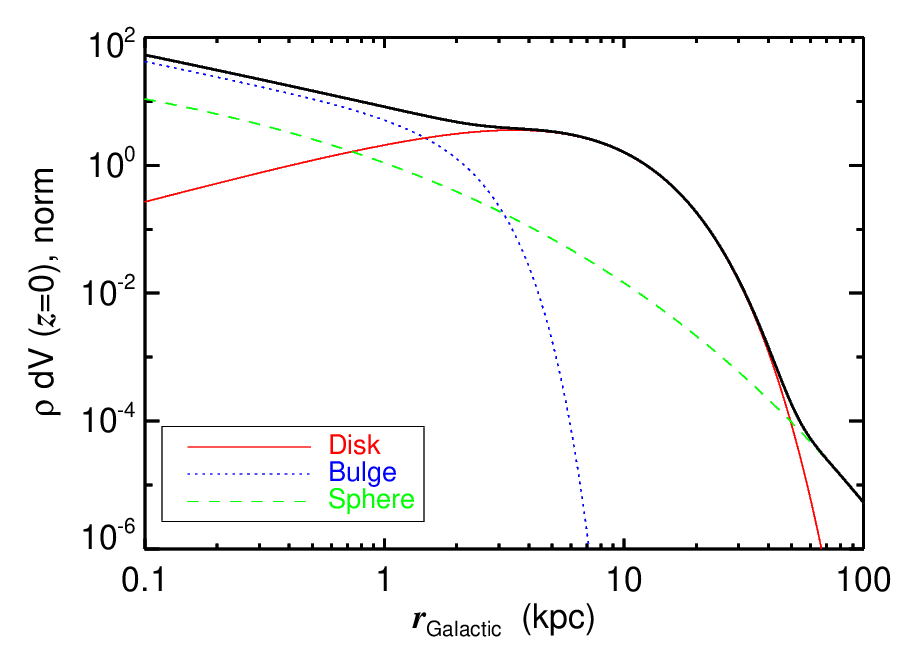} 
     \includegraphics[width=8.5cm]{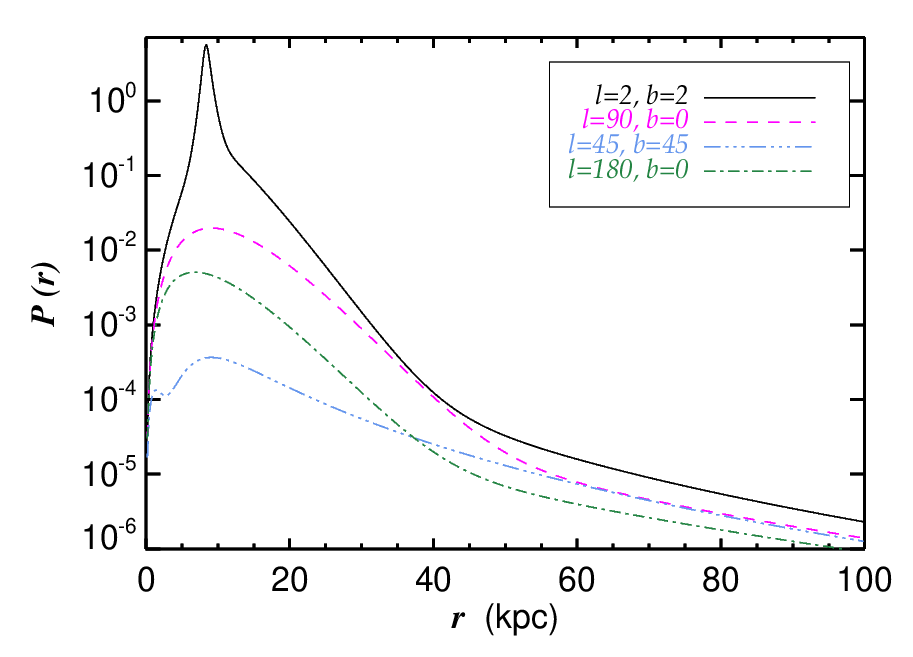}
     \caption{({\em Left:}) Galactic model showing the disc, bulge and spheroid (halo) decomposition (space density multiplied by the corresponding volume element), according to \citet{grimm02}. The model is 3-dimensional, but only a slice at $z$\,=\,0 is shown for clarity. The x-axis represents distance in the plane from the Galactic centre. ({\em Right:}) Unnormalised priors of distance $r$ from Sun resulting from the Galactic model, along several representative lines-of-sight denoted by their Galactic longitude ($l$) and latitude ($b$) expressed in degrees.}
 \label{fig:grimm02}    
 \end{figure*}

The above model is three-dimensional. The prior of relevance for this work is the one-dimensional $P(r, l, b)$, with $r$ representing distance from the Sun, and is determined by multiplying the densities by the corresponding volume element and subsequently transforming from Galacto-centric to Solar-centric coordinates. Mass density is taken to follow space density. Fig\,\ref{fig:grimm02} shows a slice through the space density model and examples of the position ($l, b$) dependent prior along several representative lines-of-sight.

\begin{table*}
  \caption{Table of Galactic and Solar constants\label{tab:constants}}
  \begin{tabular}{lccr}
    \hline
    \hline
    Constant & & Value & Units\\
    \hline
    \Rsun & Solar distance from Galactic centre & 8.34\,\p\,0.16 & kpc\\
    \zsun & Solar height above Galactic plane & 0.024 & kpc\\
    $(U_\odot, V_\odot, W_\odot)$ & Solar motion & $(10.7 \pm 1.8, 15.6 \pm 6.8, 8.9 \pm 0.9)$ & km\,s$^{-1}$\\
    $\Theta_\odot$ & Galactic rotation speed at \Rsun & 240\,\p\,8 & km\,s$^{-1}$\\
$q$      & oblateness of bulge        & 0.6     & \\
$\gamma$ & --                         & 1.8     &\\
$R_e$    & scale length of halo   & 2.8 &kpc \\
$b_h$      & --                         & 7.669   &\\
$r_0$    & scale length of bulge      & 1 &kpc   \\
$r_t$    & truncation radius of bulge & 1.9 &kpc \\
$r_d$    & scale length of disc       & 3.5 &kpc \\
$r_z$    & vertical scale of disc     & 0.41&kpc \\
$r_m$    & inner disc cut-off         & 6.5 &kpc\\
$\rho^0_d$ & disc normalisation factor       & 2.79 & \Msun\,pc$^{-3}$ \\
$\rho^0_b$ & bulge normalisation factor      & 1.19 & \Msun\,pc$^{-3}$ \\
$\rho^0_h$ & halo normalisation factor      & 22.38 & \Msun\,pc$^{-3}$ \\
$M_b$  & Integrated bulge mass & 1.3 & 10$^{10}$\,\Msun\\ 
    & Density ratio Disc:Bulge:Halo & 2:1:0.8\\
    \hline
  \end{tabular}
  ~\par
Solar constants taken from \citet{reid14}, \citet{allen}, and the Milky Way structural constants from \citet{grimm02}.
\end{table*}

\subsection{Distance estimate comparisons}

\begin{figure}
 \centering
     \includegraphics[width=85mm,angle=0]{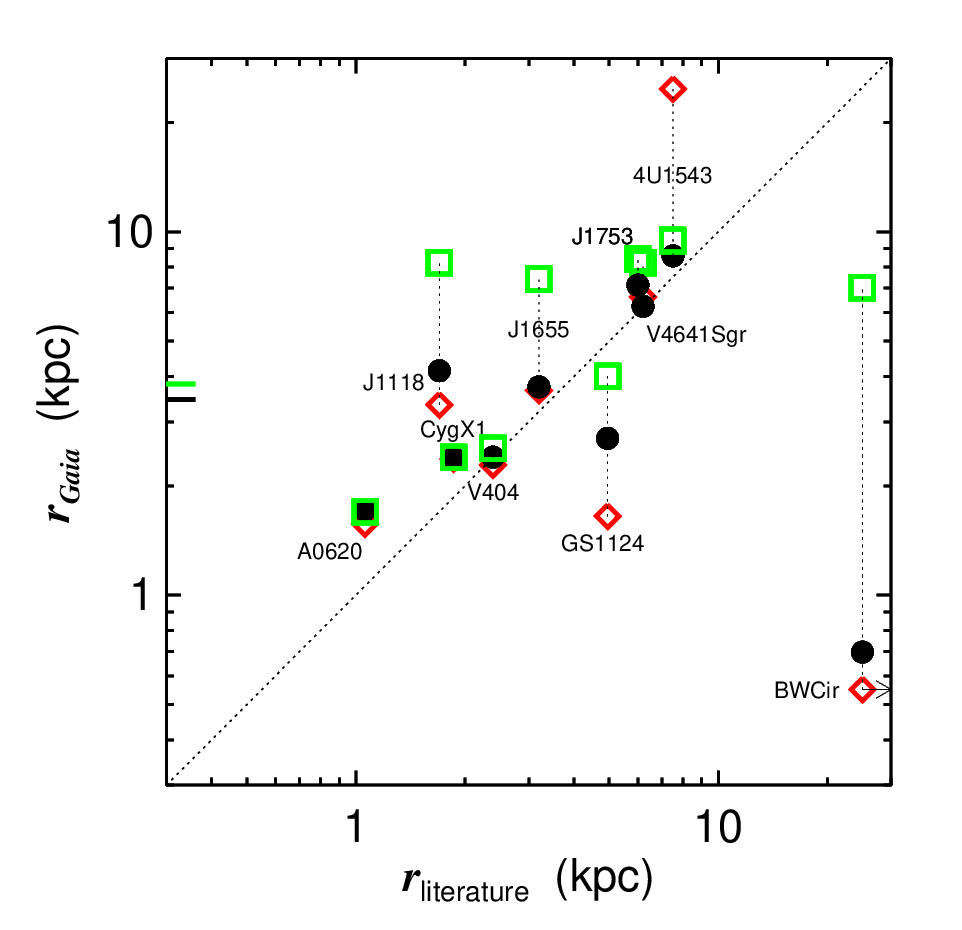} 
     \caption{Comparison of {\em Gaia} DR2 distances with previous estimates for our black hole binary targets, where known. The dotted line represents equality between the two axes. The symbols represent: $r_{\rm exp}$ as black filled circles (as in Fig.\,\ref{fig:distancecomparison}), $r_{\rm inv}$ as empty red diamonds, and $r_{\rm MW}$ as empty green squares. These three estimates are connected with a dotted line for each source. The long dashes on the left-hand vertical axis denote the most likely value of $r_{\rm exp}$ (black) and $r_{\rm MW}$ (green) for MAXI\,J1820+070, a new XRB that lacks $r_{\rm lit}$ estimates.
 \label{fig:distancecomparison2}}
 \end{figure}

Fig.\,\ref{fig:distancecomparison2} displays a comparison between the four distance estimates discussed in the main body of text: $r_{\rm inv}$, $r_{\rm exp}$ and $r_{\rm MW}$ are the three DR2 based estimates, and these are compared to the literature estimates $r_{\rm lit}$.

\newpage
\subsection{Distance and peculiar velocity posterior distributions}

\begin{figure*}
\centering
  \includegraphics[width=8.5cm]{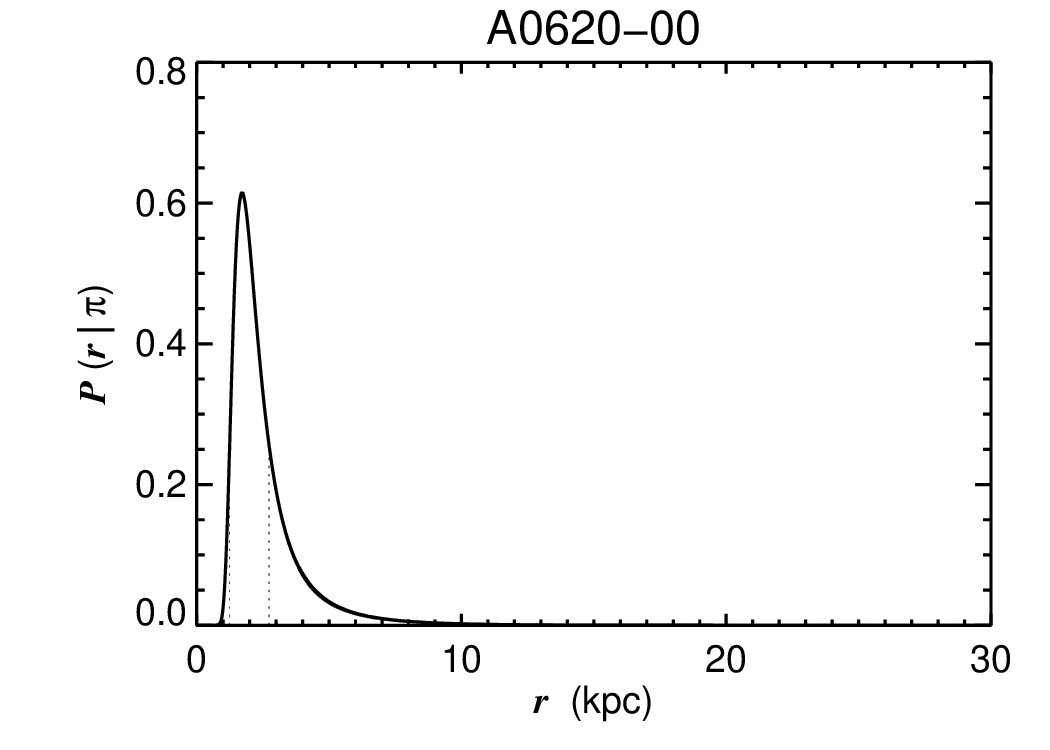}
  \includegraphics[width=8.5cm]{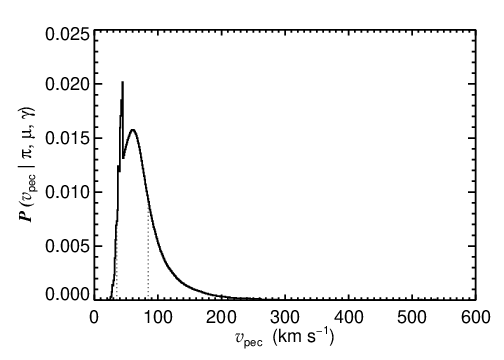}
  \caption{A\,0620--00. {\em (Left)} Distance $r_{\rm exp}$ and {\em (Right)} peculiar velocity $\upsilon_{\rm pec}$ posterior distributions. The dotted vertical lines denote the quoted confidence intervals.
    \label{fig:posteriorfirst}}
\end{figure*}

\begin{figure*}
\centering
  \includegraphics[width=8.5cm]{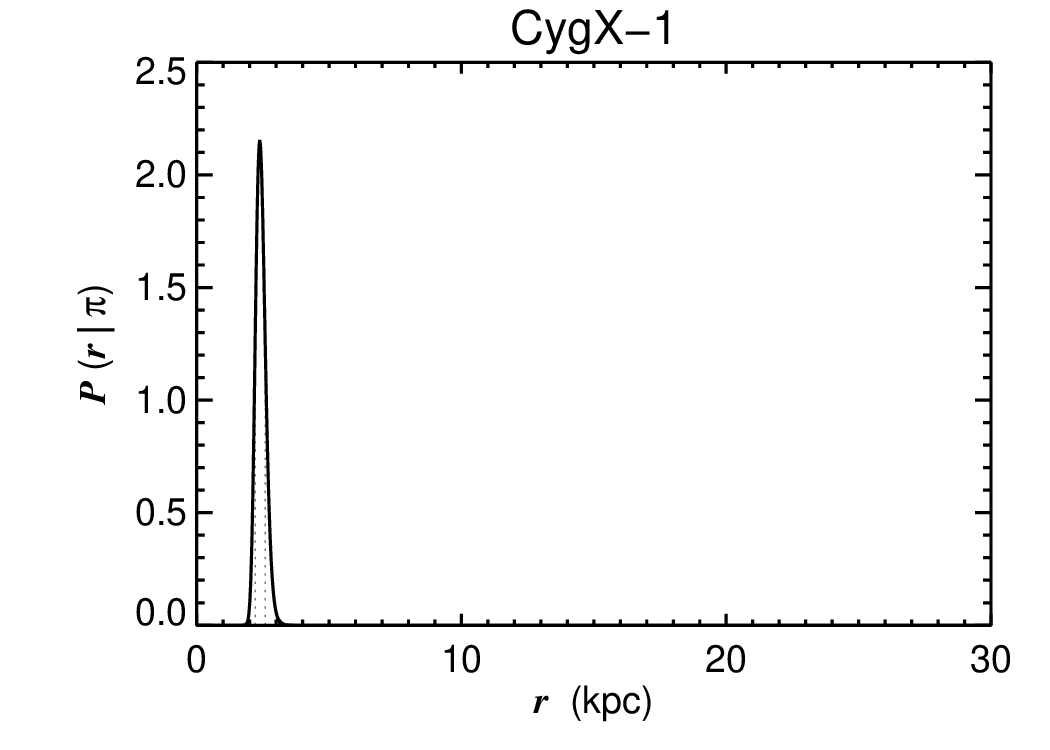}
  \includegraphics[width=8.5cm]{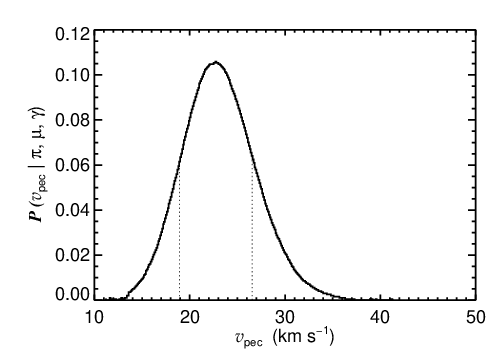}
  \caption{Cyg\,X--1. {\em (Left)} Distance $r_{\rm exp}$ and {\em (Right)} peculiar velocity $\upsilon_{\rm pec}$ posterior distributions. The dotted vertical lines denote the quoted confidence intervals. 
    \label{fig:posteriorlast}}
\end{figure*}

For each source from Table\,\ref{tab:sourceswithastrometry}, the figures \ref{fig:posteriorfirst}--\ref{fig:posteriorlast} show the $r_{\rm exp}$ distance posterior and the \vpec\ posterior, unless otherwise stated.

\newpage

\begin{figure*}
\centering
  \includegraphics[width=8.5cm]{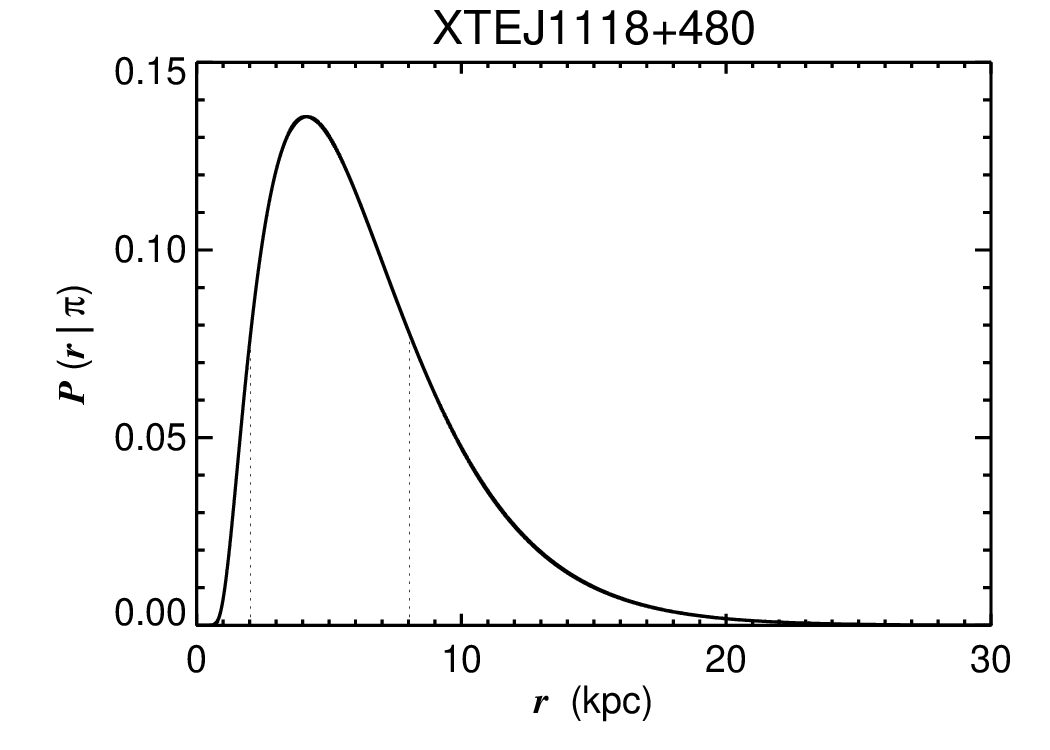}
  \includegraphics[width=8.5cm]{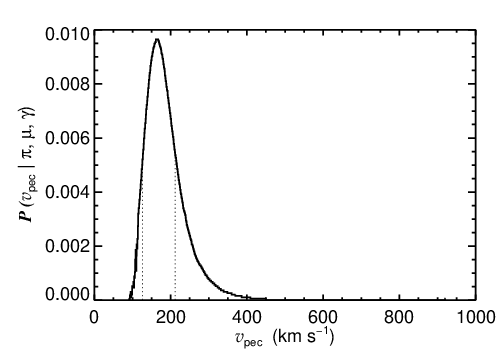}
  \caption{XTE\,J1118+480. {\em (Left)} Distance $r_{\rm exp}$ and {\em (Right)} peculiar velocity $\upsilon_{\rm pec}$ posterior distributions. The dotted vertical lines denote the quoted confidence intervals. 
    In this case, the fractional DR2 parallax error is particularly poor, $\frac{\sigma_\pi}{\pi}$\,$>$\,1, so $r_{\rm lit}$ is combined with the DR2 proper motions and published $\gamma$ velocity to determine \vpec.}
\end{figure*}
\newpage
    
\begin{figure*}
\centering
  \includegraphics[width=8.5cm]{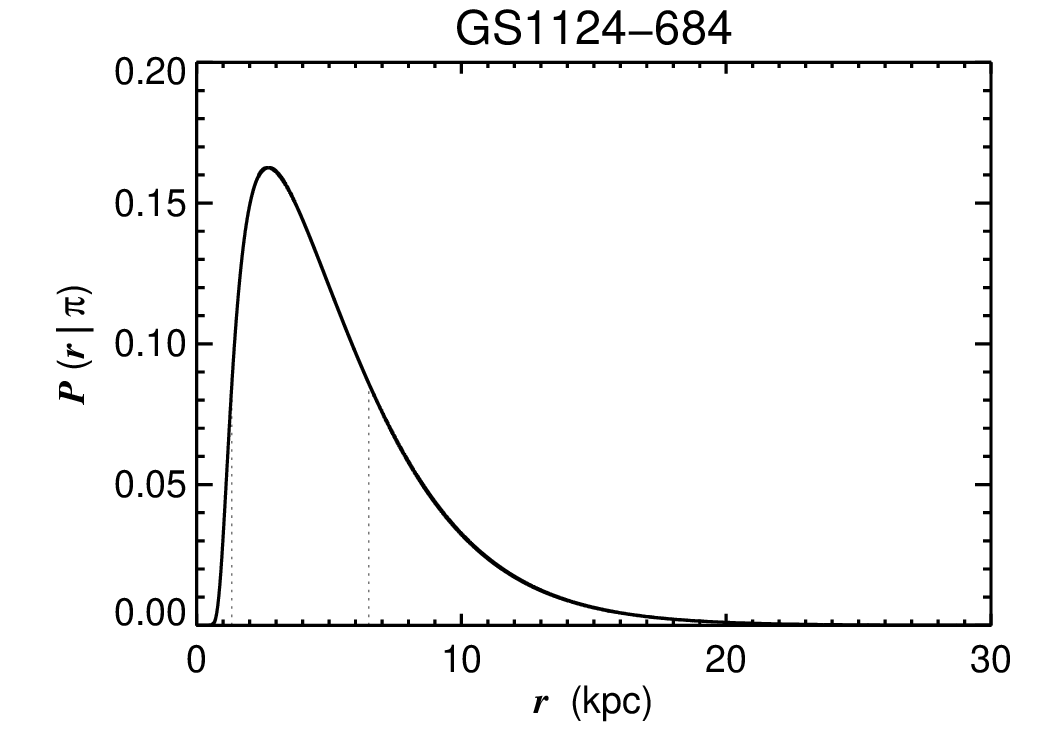}
  \includegraphics[width=8.5cm]{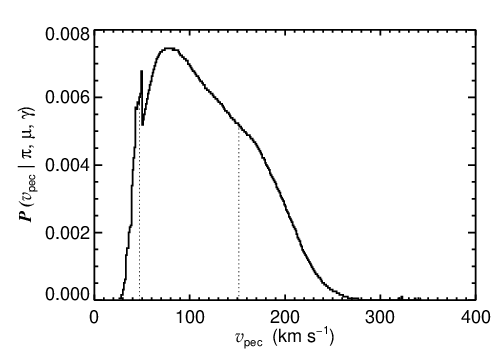}
  \caption{GS\,1124--684. {\em (Left)} Distance $r_{\rm exp}$ and {\em (Right)} peculiar velocity $\upsilon_{\rm pec}$ posterior distributions. The dotted vertical lines denote the quoted confidence intervals. 
  }  
\end{figure*}
\newpage
    
\begin{figure*}
\centering
  \includegraphics[width=8.5cm]{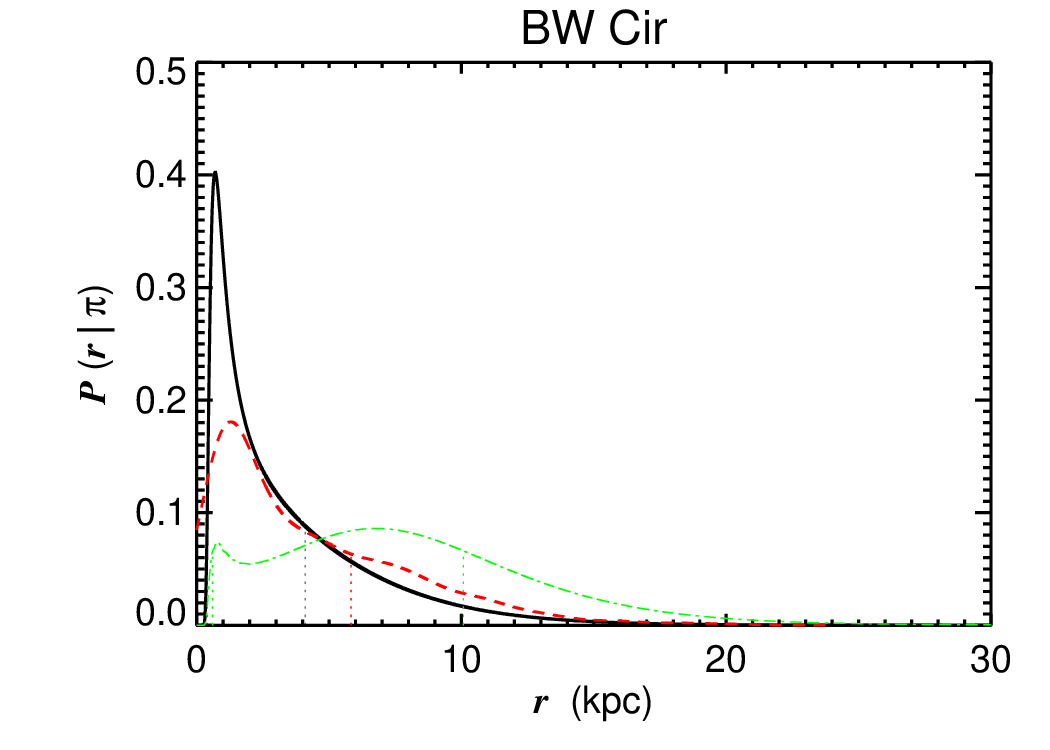}
  \includegraphics[width=8.5cm]{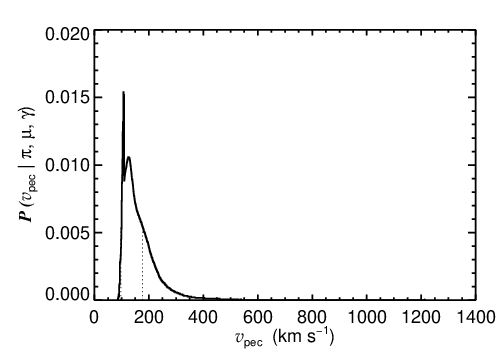}
  \caption{BW\,Cir. {\em (Left)} Distance $r_{\rm exp}$ and {\em (Right)} peculiar velocity $\upsilon_{\rm pec}$ posterior distributions. The dotted vertical lines denote the quoted confidence intervals. For this source, the distance plot additionally shows the effect on the posterior when including high potential covariances between the DR2 astrometric fit parameters for this source (red dashed curve), as well as the Milky Way posterior (green dot-dashed curve). Both demonstrate the unlikelihood of the source lying at distances of $\gtsim$\,25\,kpc. 
  }
\end{figure*}
\newpage
    
\begin{figure*}
\centering
  \includegraphics[width=8.5cm]{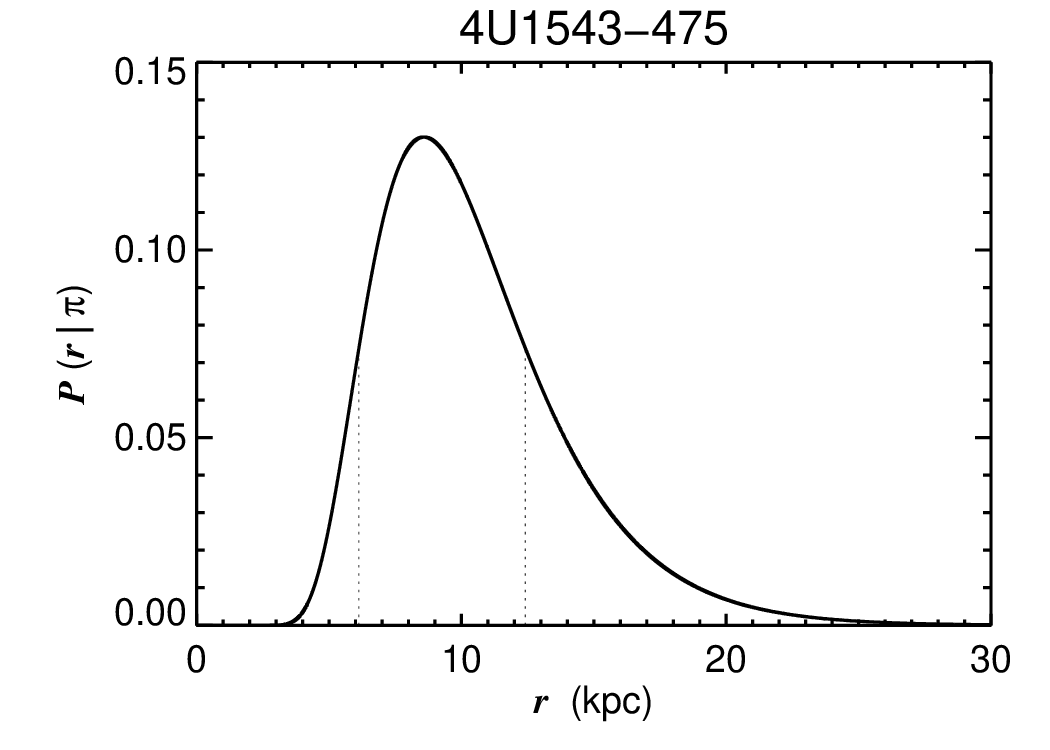}
  \includegraphics[width=8.5cm]{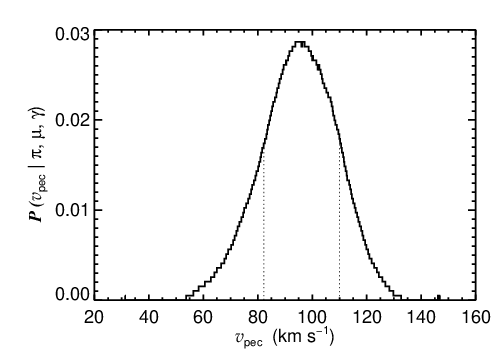}
  \caption{4U\,1543--475. {\em (Left)} Distance $r_{\rm exp}$ and {\em (Right)} peculiar velocity $\upsilon_{\rm pec}$ posterior distributions. The dotted vertical lines denote the quoted confidence intervals. 
    In this case, the fractional DR2 parallax error is particularly poor, $\frac{\sigma_\pi}{\pi}$\,$>$\,1, so $r_{\rm lit}$ is combined with the DR2 proper motions and published $\gamma$ velocity to determine \vpec.} 
\end{figure*}
\newpage
    
\begin{figure*}
\centering
  \includegraphics[width=8.5cm]{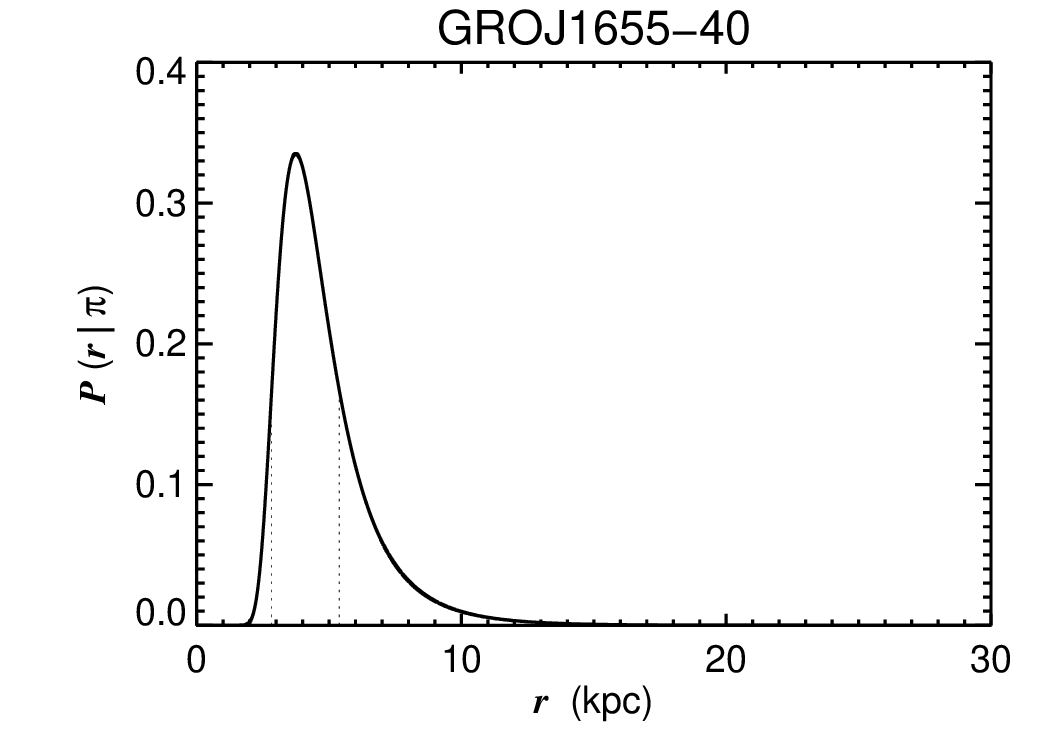}
  \includegraphics[width=8.5cm]{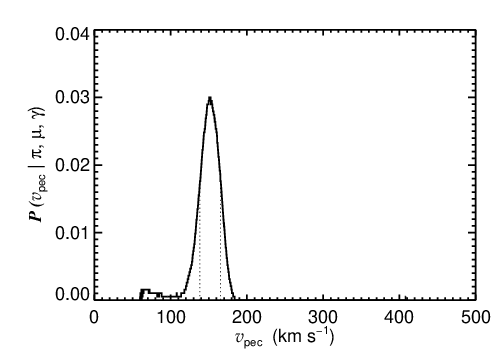}
  \caption{GRO\,J1655-40. {\em (Left)} Distance $r_{\rm exp}$ and {\em (Right)} peculiar velocity $\upsilon_{\rm pec}$ posterior distributions. The dotted vertical lines denote the quoted confidence intervals.
  } 
\end{figure*}
\newpage
    
\begin{figure*}
\centering
  \includegraphics[width=8.5cm]{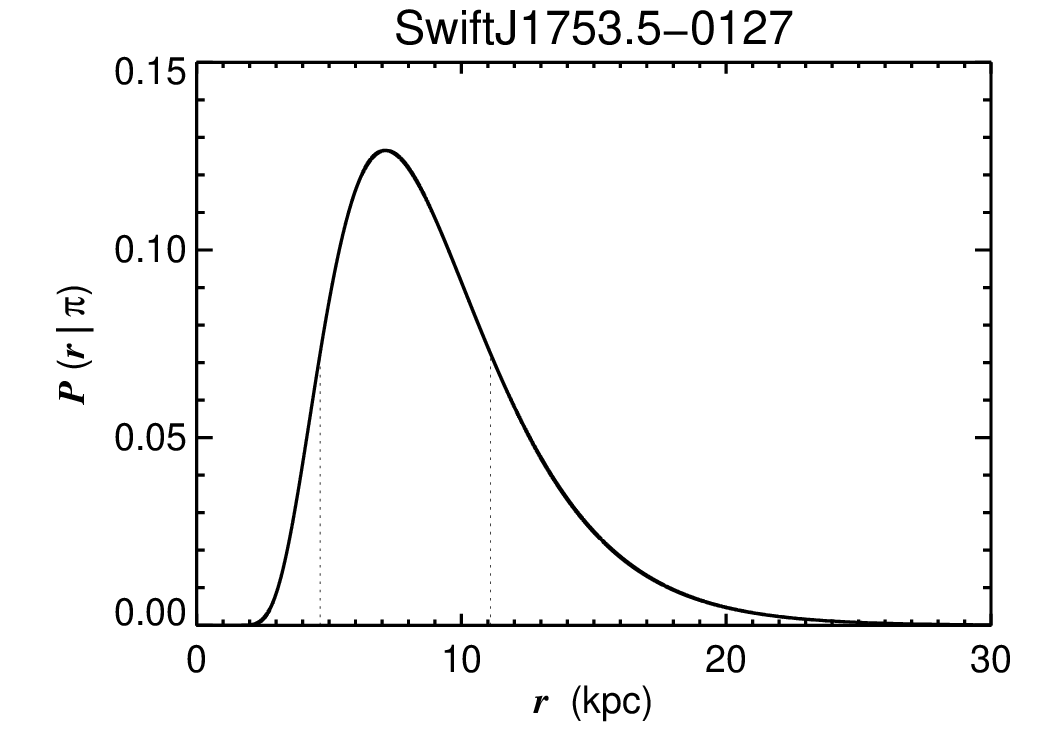}
  \includegraphics[width=8.5cm]{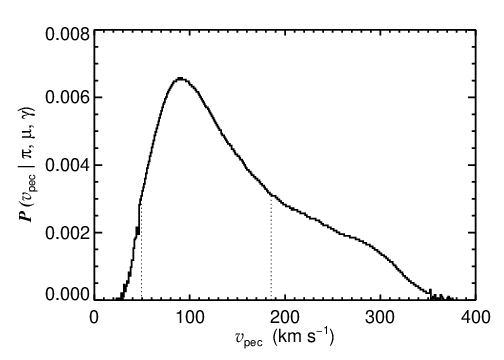}
  \caption{Swift\,J1753.5--0127. {\em (Left)} Distance $r_{\rm exp}$ and {\em (Right)} peculiar velocity $\upsilon_{\rm pec}$ posterior d
    In this case, the fractional DR2 parallax error is particularly poor, $|\frac{\sigma_\pi}{\pi}|$\,$>$\,1, so $r_{\rm lit}$ is combined with the DR2 proper motions and published $\gamma$ velocity to determine \vpec.}
\end{figure*}
\newpage
    
\begin{figure*}
\centering
  \includegraphics[width=8.5cm]{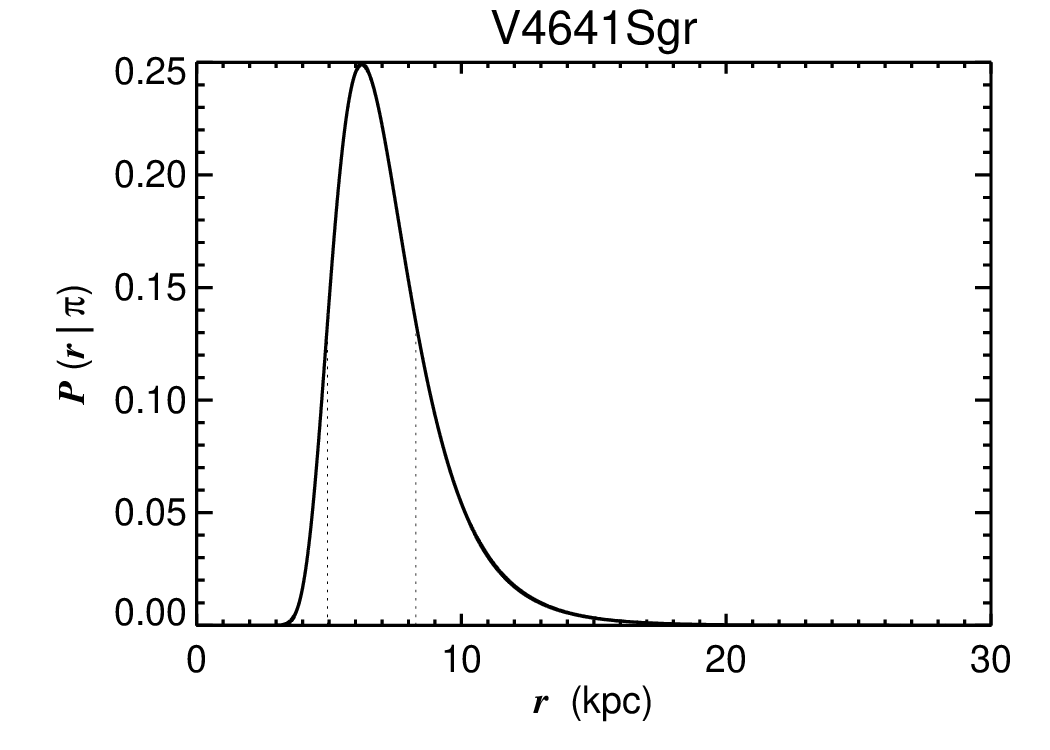}
  \includegraphics[width=8.5cm]{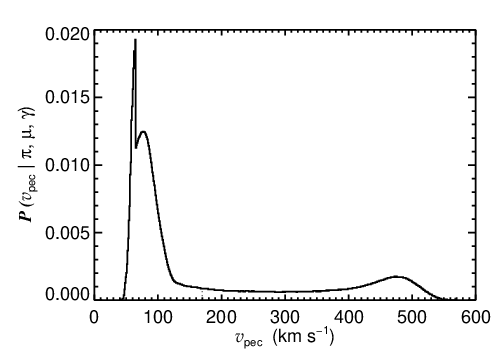}
  \caption{V4641\,Sgr. {\em (Left)} Distance $r_{\rm exp}$ and {\em (Right)} peculiar velocity $\upsilon_{\rm pec}$ posterior distributions. The dotted vertical lines denote the quoted confidence intervals. 
  }
\end{figure*}
\newpage
    
\begin{figure*}
\centering
  \includegraphics[width=8.5cm]{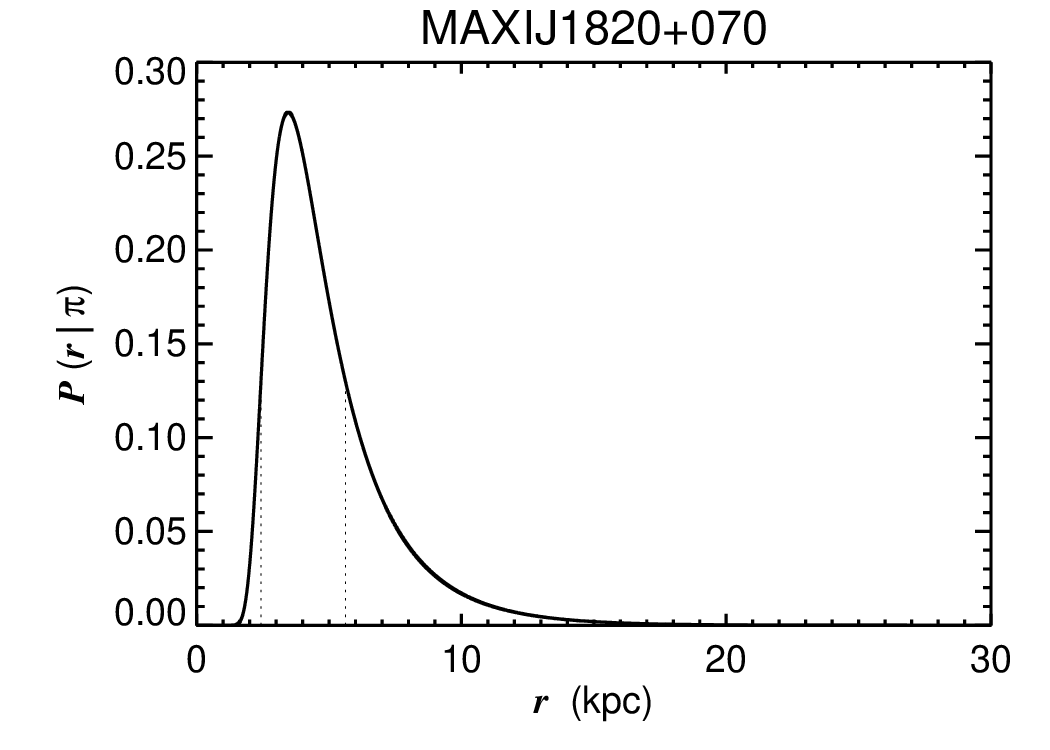}
  \includegraphics[width=8.5cm]{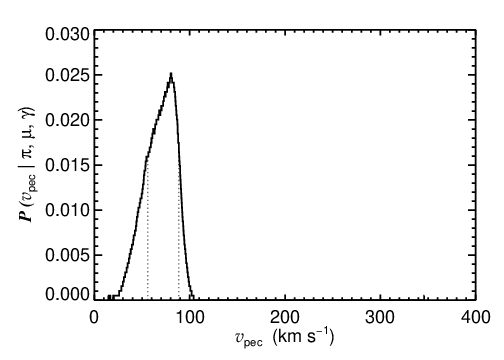}
  \caption{MAXI\,J1820+070. {\em (Left)} Distance $r_{\rm exp}$ and {\em (Right)} peculiar velocity $\upsilon_{\rm pec}$ posterior distributions. The dotted vertical lines denote the quoted confidence intervals. 
    For this source, the systemic radial velocity is currently unknown, and is assumed to be $\gamma$\,=\,0\,km\,s$^{-1}$.}
\end{figure*}
\newpage
    
\begin{figure*}
\centering
  \includegraphics[width=8.5cm]{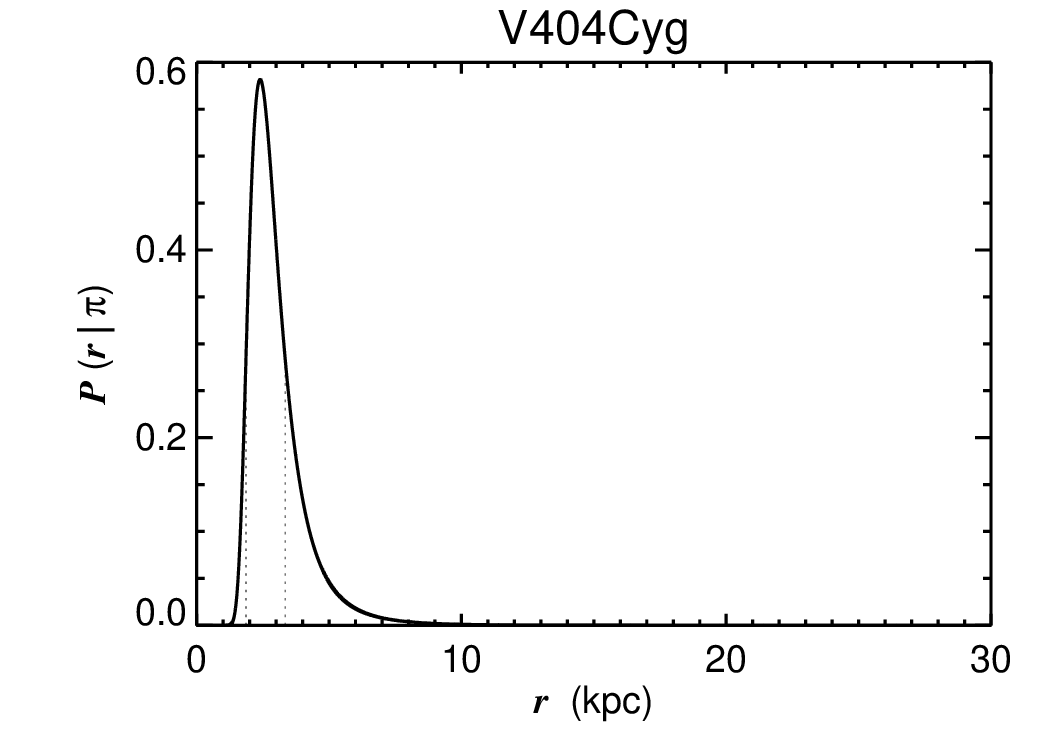}
  \includegraphics[width=8.5cm]{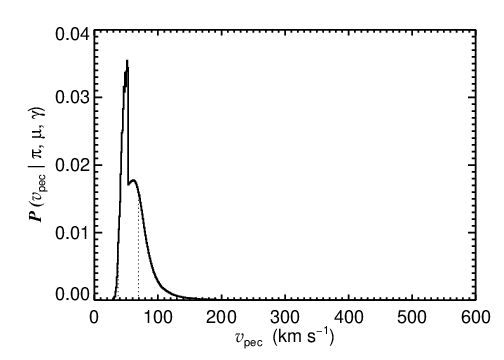}
  \caption{V404\,Cyg. {\em (Left)} Distance $r_{\rm exp}$ and {\em (Right)} peculiar velocity $\upsilon_{\rm pec}$ posterior distributions. The dotted vertical lines denote the quoted confidence intervals. 
  }
\end{figure*}
\newpage
    
\end{document}